\documentclass[aps,prl,twocolumn,superscriptaddress,showpacs]{revtex4-2}
\usepackage{amsmath, amssymb,wasysym,graphicx}
\usepackage{appendix}
\usepackage[ruled]{algorithm2e}
\usepackage{lineno}

\usepackage[utf8]{inputenc}
\usepackage[T1]{fontenc}
\usepackage{xcolor}

\usepackage{diagbox}
\usepackage{tabularx}

\IfFileExists{newtxtext.sty}
{\usepackage{newtxtext,newtxmath}}
{\IfFileExists{stix.sty}
{\usepackage{stix}}
{\IfFileExists{mathptmx.sty}
{\usepackage{mathptmx}}{} } }

\usepackage{textcomp}

\usepackage{bm}

\usepackage{booktabs,siunitx}

\pdfoutput=1
\usepackage{color}
\definecolor{LinkColor}{rgb}{0.256,0.439,0.588}
\usepackage{hyperref}
\hypersetup{
colorlinks=true,
citecolor=LinkColor,
linkcolor=LinkColor,
urlcolor=LinkColor
}

\usepackage{multirow}

\begin{document}
\title{Universal term of Entanglement Entropy in the $\pi$-flux Hubbard model}


\author{Yuan Da Liao}
\email{ydliao@hku.hk}
\affiliation{Department of Physics and HK Institute of Quantum Science \& Technology, The University of Hong Kong, Pokfulam Road, Hong Kong SAR, China}

\date{\today}

\begin{abstract}
Researchers in physical science aim to uncover universal features in strongly interacting many-body systems, often hidden in complicated observables like entanglement entropy (EE).  The non-local nature of EE makes it challenging to compute numerically, necessitating the development of an unbiased and convenient algorithm. In this paper, we use quantum Monte Carlo to reveal that the coefficient of variation in direct EE calculations increases exponentially with system size, leading to inaccuracies. To address this issue, we develop a power incremental algorithm and a technique for straightforwardly calculating the universal term of EE, successfully evaluating the EE of a 2D Hubbard model. Our numerical results demonstrate the consistency of the universal coefficient of EE from sharp corners at the Gross-Neveu quantum critical point and for free Dirac fermions. Our method can also be applied to other unstable observables, such as partition functions, entanglement spectra, and negativity, thereby fostering computational and theoretical progress.
\end{abstract}

\maketitle

{\it Introduction.---}
Physical science researchers are always seeking for the universal properties in correlated many-body systems~\cite{klebanovRenyi2012,altmanUniversal2015}. These properties can be difficult to detect because they are often hidden within complicated observables that have a non-local nature, which means that it is hard to measure them experimentally or to even evaluate them numerically. One example of such an observable is the entanglement entropy (EE), that is a measure of the information entanglement of the boundary of subsystems~\cite{eisertColloquium2010,casiniUniversal2007}.
The EE generally shows a finite-size scaling relation, and the corresponding scaling coefficients, that are related to the low-energy physics and subsystem geometry, can manifest such important universal features.

For instance, in one-dimensional critical systems, EE scales logarithmically with respect to the subsystem size and has a universal leading coefficient proportional to the central charge~\cite{vidalEntanglement2003,calabreseEntanglement2004,korepinUniversality2004a}. 
In two and higher dimensions, although the theory of EE is not completely known, there is a consensus that EE grows proportionally to the area of the subsystem boundary for both critical and non-critical systems~\cite{eisertColloquium2010}, known as "area law". 
Interestingly, for critical systems, a subleading universal and geometry-dependent logarithmic term will be displayed~\cite{casiniUniversal2007,songEntanglement2011}, as is the case for free Dirac cones~\cite{helmesUniversal2016} and quantum critical points (QCPs)~\cite{demidioUniversal2022,daliaoteaching2023}.
Furthermore, due to the interaction of Goldstone modes and the restoration of symmetry in a limited space, a universal subleading logarithmic term will arise~\cite{metlitski2015entanglement}. The coefficient of the universal term is equal to half the number of Goldstone modes and applies to both spin and fermion models with continuous spontaneous symmetry breaking~\cite{songEntanglement2011,kallinAnomalies2011,emidioEntanglement2020,zhaoMeasuring2022,
demidioUniversal2022,daliaoteaching2023,panComputing2023,dengImproved2023}.
All these intriguing facts represent that the EE can be a potent tool for revealing the universal features of critical systems.

\begin{figure}[htp!]
\includegraphics[width=1\columnwidth]{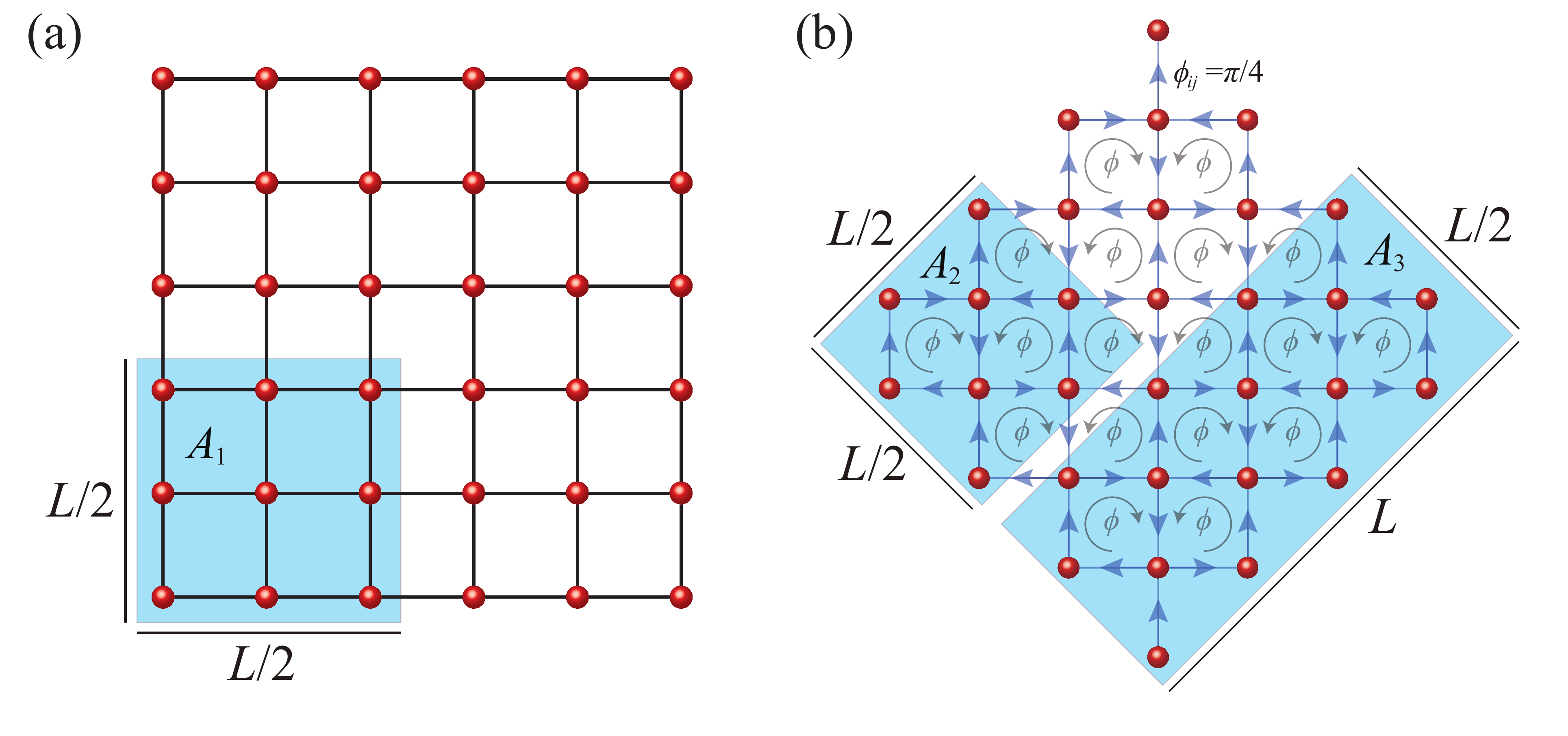}
\caption{Entanglement region shapes. (a) A $L/2\times L/2$ square-shaped region $A_1$ with $L^2/4$ sites, featuring four $90^\circ$ sharp corners on a square lattice. (b) A $L/2\times L/2$ square-shaped region $A_2$ with $L^2/2$ sites, exhibiting four $90^\circ$ sharp corners, and a $L \times L/2$ region $A_3$ with $L^2$ sites, showcasing a smooth boundary on a square lattice. The boundary length $l=2L$ for both $A_2$ and $A_3$.}
\label{fig1}
\end{figure}

The theoretical explanation of the universal term of the n-th R\'enyi EE beyond the area law scaling for free Dirac fermions in 2D has become apparently clear~\cite{klebanovRenyi2012,helmesUniversal2016}.
However, for interacting fermions, the complete theoretical understanding is not yet well-established.
Therefore, a viable alternative to extract the universal term of the EE for interacting fermions is the use of numerical tools.  
The auxiliary-field determinant quantum Monte Carlo (DQMC) simulation is an appropriate and unbiased numerical tool to study the properties of interacting fermions for cases without the sign problem.
In recent years, significant algorithmic advances have been made for EE.
Grover's pioneering work~\cite{groverEntanglement2013} presented a seminal numerical definition of the 2nd R\'enyi EE $S_2^A$ for interacting fermions, here $A$ is the entangled region we choose, as shown in Fig.~\ref{fig1}. 
This definition is given by $e^{-S_2^A}=\langle \operatorname{det} g^{s_1, s_2}_A \rangle$, and a matrix $g^{s_1, s_2}_A$, that relates to two independent replica configurations of the auxiliary field $\{s_1\}$ and $\{s_2\}$, is introduced. 
However, direct statistics of $\operatorname{det}  g^{s_1, s_2}_A$ suffer from severe instability at slightly stronger coupling and slightly larger subsystem sizes~\cite{panComputing2023}. This instability hinders the precise determination of the EE.
 
Over the past decade, many attempts have been made to solve this problem~\cite{broeckerRenyi2014,wangRenyi2014,demidioUniversal2022}, but none have achieved a perfect solution. 
Methods used a few years ago involved introducing an extra entangling subsystem that effectively enlarged the total system size, which offered better controlled statistical errors but became cumbersome in practical simulations~\cite{broeckerRenyi2014,wangRenyi2014}. 
A recent incremental method~\cite{demidioUniversal2022} introduces an additional configuration space, leading to an increased computational burden.
A more critical issue is that the estimator employed in Ref.~\cite{demidioUniversal2022} failed to revert to the original definition of EE proposed by Grover~\cite{groverEntanglement2013} when the total number of incremental processes was reduced to one. This discrepancy led to bias in the results. For a detailed discussion of this issue, please refer to the Supplementary Materials (SM).
Consequently, the total number of incremental processes generally needs to be quite large, but it can only be estimated empirically rather than being quantitatively determined.
These limitations render the previous methods challenging to apply in practice, and there is an urgent need for the development of an improved method.
Moreover, despite considerable research, the underlying cause of the instability encountered during the direct calculation of EE remains a subject of ongoing investigation and has not yet been fully elucidated.

In this paper, we explicitly demonstrate for the first time the true underlying cause of instability in the direct calculation of EE, and develop a power incremental method to resolve this problem.
To extract the universal feature of EE more effectively, we further propose a straightforward technique to obtain the universal term of EE directly. This novel approach not only can calculate the universal term of EE arising from corners in the symmetry spontaneous broken (SSB) phase, but also circumvents the need for complicated function fitting with limited data points and reduces the computational cost.
Moreover, our method can be readily extended to calculate other physical observables such as partition functions, entanglement spectra, and negativity of correlated boson/spin and fermion systems, and other general functionals of determinants of Green functions in interacting fermions.

{\it Methods details and numerical results---}
We firstly investigate the $0$-flux Hubbard model to showcase the effectiveness and convenience of our power incremental method. 
This model is well investigated, and the ground state is the antiferromagnet at finite interaction strength $U$~\cite{qinHubbard2022}.
We employ our power incremental method to calculate the $S_2^{A}$ at $U=10$ with square entanglement region $A_1$ as shown in Fig.~\ref{fig1} (a) .

\begin{figure}[htp!]
\includegraphics[width=1\columnwidth]{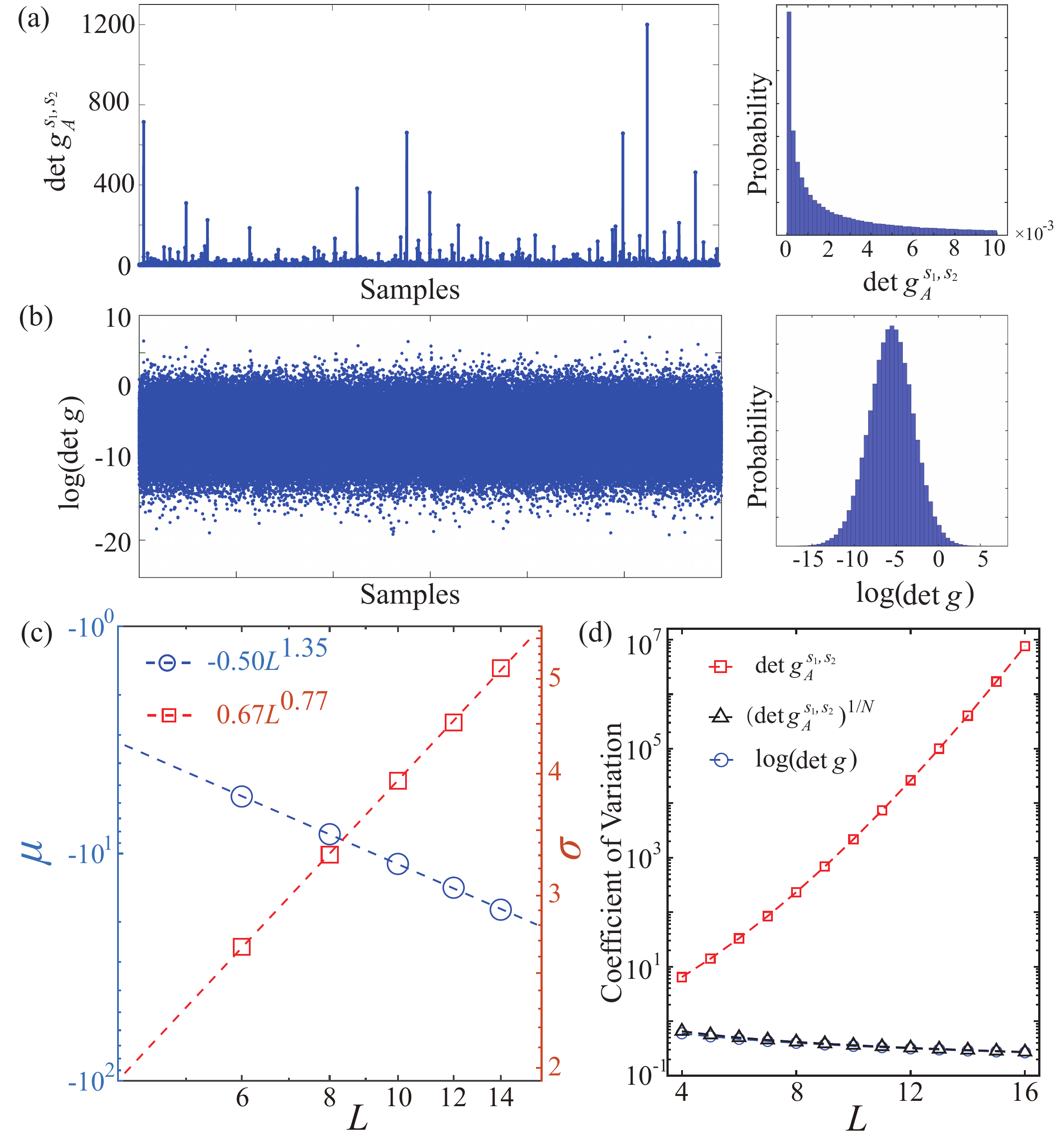}
\caption{(a) and (b) show the sampling distributions and normalized  histograms of $\operatorname{det} g_A^{s_1, s_2}$ and $\log (\operatorname{det} g)$ with $U=10$ and $L=6$. We notice that the statistical distribution of $\log (\operatorname{det} g)$  conforms to the normal form $\mathcal{N}(\mu,\sigma)$, while  $\operatorname{det} g_A^{s_1, s_2}$ apparently not. (c) The log-log plot of $\mu$ (left) and $\sigma$ (right) as function of $L$ with $U=10$, respectively. The data can be fit with a perfect scaling relation $\mu(L) = -0.50 L^{1.35}$ and  $\sigma(L) = 0.67 L^{0.77}$. Here we estimate $\mu$ and $\sigma$ with approximate $10^6$ samples.(d) The log plot of coefficient of variation for there estimates. It's clear that $\text{CV}[\operatorname{det} g_A^{s_1, s_2}]$ exponentially explodes. While $\text{CV}[(\operatorname{det} g_A^{s_1, s_2})^{1/N}]$ and $\text{CV}[\log (\operatorname{det} g)]$ decay respect to $L$. }
\label{fig2}
\end{figure}

The numerical definition of $S_2^A$ proposed by Grover~\cite{groverEntanglement2013} is 
\begin{equation}\label{eq:1}
e^{-S_2^A}=\frac{\sum_{\left\{s_1\right\},\left\{s_2\right\}} W_{s_1,s_2} \operatorname{det} g_A^{s_1, s_2}}{\sum_{\left\{s_1\right\},\left\{s_2\right\}} W_{s_1,s_2}},
\end{equation}
where $W_{s_1,s_2}$ represents the unnormalized weight, Grover matrix, defined as $g_A^{s_1,s_2}= G_A^{s_1} G_A^{s_2}+\left(\mathbb{I}-G_A^{s_1}\right)\left(\mathbb{I}-G_A^{s_2}\right) $, is a functional of Green's functions $G_A^{s_1}$ and $G_A^{s_2}$ of two independent replica.
The direct measurement of $\langle \operatorname{det} g_A^{s_1, s_2} \rangle$ becomes extremely unstable at slightly stronger coupling and larger system sizes. 
This instability appears to stem from a phenomenon in which the distribution of $\operatorname{det} g_A^{s_1, s_2}$ becomes highly skewed, exhibiting occasional sharp peaks, as shown in Fig.\ref{fig2} (a). 
These rare spikes significantly influence and are argued to dominate the expectation value of $\operatorname{det} g_A^{s_1, s_2}$ in Refs.\cite{demidioUniversal2022,panComputing2023}.
However, we believe that the true cause of such instability is the log-normal sampling distribution of $\operatorname{det} g_A^{s_1, s_2}$, which implies that the expectation value of the EE is not dominated by rare spikes, but rather all samplings contribute significantly to it.

As shown in Fig.~\ref{fig2} (b), we initially observed that the logarithm of $\operatorname{det} g_A^{s_1, s_2}$, denoted as $\log(\operatorname{det} g)$, conforms very well to a normal distribution $\mathcal{N}(\mu,\sigma^2)$. This implies that the distribution of $\operatorname{det} g_A^{s_1, s_2}$ exhibits a log-normal behavior.
We harness normal distribution functions to fit the mean $\mu$ and standard deviation $\sigma$ of $\log(\operatorname{det} g)$ for different $L$ with approximate $10^6$ samples.
As shown in Fig.~\ref{fig2} (c), $\mu$ and $\sigma$ follow power functions of the perfect form $\mu\sim L^{\gamma_\mu}$ and $\sigma\sim L^{\gamma_\sigma}$, respectively.
These data imply that the coefficient of variation of $\log(\operatorname{det} g)$, defined as $\text{CV}[\log(\operatorname{det} g)] = \sigma/|\mu|$, decays as a power relative to $L$.
However, based on our numerical observation, we know $\text{CV}[\operatorname{det} g_A^{s_1, s_2}]=\sqrt{e^{\sigma^2}-1}$ that would disastrously exponentially grow as $L$ increasing.
According to the central limit theorem, the exponentially increased computational burden makes it infeasible to estimate the exponential observable $e^{-S_2^A}$ with high accuracy. This is the true underlying reason why previous algorithms were unsuccessful in directly calculating the EE based on Eq.~\eqref{eq:1}.

Fortunately, the $1/N$-th power of $\operatorname{det} g_A^{s_1, s_2}$ has reduced coefficient of variation $\text{CV}[(\operatorname{det} g_A^{s_1, s_2})^{1/N}] = \sqrt{e^{\sigma^2/N^2}-1} $ based on our numerical observation.
If we permit $N$ to appropriately scale as $L$, like $N \sim L^{\gamma}$ and making sure $\gamma > \gamma_\sigma$ in our cases, $\text{CV}[(\operatorname{det} g_A^{s_1, s_2})^{1/N}]$ would decay with $L$, enabling accurate evaluation.
Here we set $N=|\mu|$ due to the fact $\gamma_\mu>\gamma_\sigma$, and plot the $\text{CV}[\operatorname{det} g_A^{s_1, s_2}]$, $\text{CV}[(\operatorname{det} g_A^{s_1, s_2})^{1/N}]$, and $\text{CV}[\log(\operatorname{det} g)]$ versus $L$ in Fig.~\ref{fig2} (d) with fitted parameters.
Evidently, $\text{CV}[\operatorname{det} g_A^{s_1, s_2}]$ explodes exponentially with $L$, while $\text{CV}[(\operatorname{det} g_A^{s_1, s_2})^{1/N}]$ and $\text{CV}[\log (\operatorname{det} g)]$ exhibit commensurate decay. This signifies the $1/N$-th power of $\operatorname{det} g_A^{s_1, s_2}$ can be computed accurately analogous to $\log(\operatorname{det} g)$.

Based on our observations and reasoning, we introduce a power incremental method. We first define
\begin{equation}
Z\left(k\right)=\sum_{\left\{s_1\right\},\left\{s_2\right\}} W_{s_1,s_2} ( \operatorname{det} g_A^{s_1, s_2})^{k/N},
\end{equation}
thus $e^{-S_2^A}=Z(N)/Z(0)$ with $Z(0) = \sum_{\left\{s_1\right\},\left\{s_2\right\}} W_{s_1,s_2}$ and  $Z({N}) = \sum_{\left\{s_1\right\},\left\{s_2\right\}} W_{s_1,s_2} \operatorname{det} g_A^{s_1, s_2} $ according to Eq.~\eqref{eq:1}.
We now proceed to rigorously rewrite Eq.~\eqref{eq:1} in a well-known yet significant incremental form~\cite{Forcrand2001}
\begin{equation}\label{eq:incremental}
 	e^{-S_2^A}=\frac{Z(1)}{Z(0)}\frac{Z(2)}{Z(1)}\cdots\frac{Z({k+1})}{Z(k)} \cdots\frac{Z({N})}{Z({N} - 1)},
\end{equation}
where the integer $k$ represents $k$-th incremental process, there are $N$ incremental process in total.
And each incremental process can be evaluated via DQMC in parallel
\begin{equation}\label{eq:4}
\frac{Z\left({k+1}\right)}{Z\left(k\right)}=\frac{\sum\limits_{\left\{s_1\right\},\left\{s_2\right\}} W_{s_1,s_2}  ( \operatorname{det} g_A^{s_1, s_2})^{k/ N} ( \operatorname{det} g_A^{s_1, s_2})^{1/N} }{\sum\limits_{\left\{s_1\right\},\left\{s_2\right\}} W_{s_1,s_2} ( \operatorname{det} g_A^{s_1, s_2})^{k/N}},
\end{equation}
where $N$ can be quantitatively determined as the ceiling integer of $|\mu|$. It's crucial to ensure that $ \operatorname{det} g_A^{s_1, s_2} $ is positive to avoid the sign problem in DQMC simulations, especially since we now include $ \operatorname{det} g_A^{s_1, s_2} $ in the updating weight. For our Hubbard model, this positivity is ensured by particle-hole (PH) symmetry. The positivity of $ W_{s_1,s_2} $ is also guaranteed by the PH symmetry.

Here we incorporate this power incremental method into projector DQMC, and we set the projection length $\beta$ to be equal to the system size $L$, with a discrete time slice of 0.1.  
The computational complexity of each incremental process remains $O(\beta L^6)$, the same as that of the standard DQMC method. 
The overall computational complexity of computing the EE is $O(\beta L^{6+\gamma_\mu})$, because of the $N \sim L^{\gamma_\mu}$ incremental steps required.
Here $\gamma_\mu=1.35$ for 0-flux Hubbard model at $U=10$. 
Certainly, determining $\gamma_\mu$ and $\gamma_\sigma$ necessitates examining the scaling of $\mu(L)$ and $\sigma(L)$. Nonetheless, this merely demands additional computations $O(\beta L^6)$. Moreover, fitting $\gamma_\mu$ and $\gamma_\sigma$ only requires a few small $L$. Hence the extra computational overhead proves negligible relative to the total cost of evaluating EE.

\begin{figure}[htp!]
\includegraphics[width=1\columnwidth]{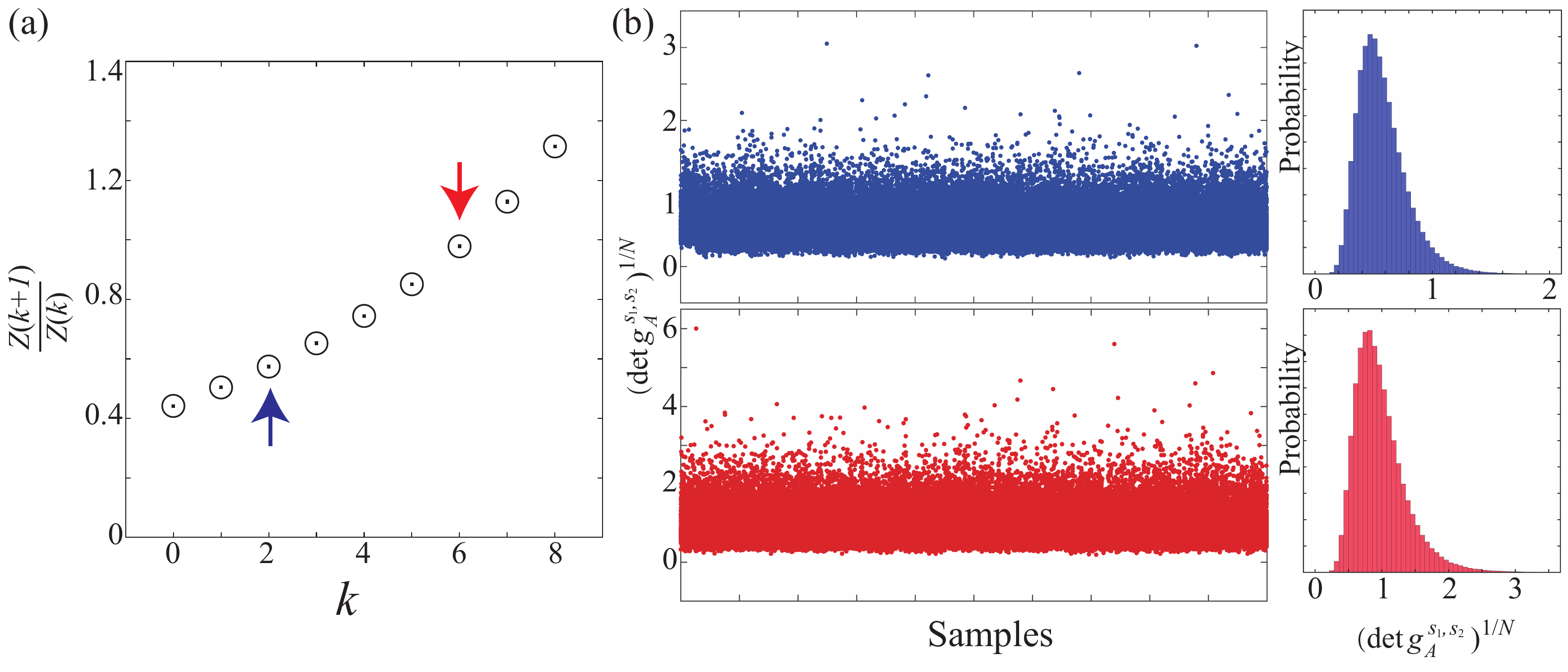}
\caption{(a). The $\frac{Z({k+1})}{Z(k)}$ as function of incremental process for $U=10$ and $L=8$. Here $N$ is $9$. We can notice that the $\frac{Z({k+1})}{Z(k)}$ of each $k$-th process can be evaluated very well, which will give rise to reliable and accurate $S_2^A$. (b) The sampling distributions and normalized histograms of new observable of our improved algorithm $(\operatorname{det} g)^{1/N}$. All pieces of incremental process generate reliable samples of $(\operatorname{det} g)^{1/N}$ that approximatively follow normal distribution and can be reasonably estimate.
}
\label{fig3}
\end{figure}

Our power incremental method allows us to evaluate the new observables $( \operatorname{det} g_A^{s_1, s_2})^{1/N} $ for each piece of parallel incremental process with small statistical errors, as illustrated in Fig.~\ref{fig3}(a). As expected, all pieces of the incremental process shown in Fig.~\ref{fig3}(b) generate reliable samples that approximately follow a normal distribution and have a finite variance.
According to Eq.~\eqref{eq:incremental}, the reliable $S_2^A$ can be obtained by multiplying all parallel expected values $\frac{Z({k+1})}{Z(k)}$ together, and the errorbar of $S_2^A$ is obtained through the propagation of uncertainty.

\begin{figure}[htp!]
\includegraphics[width=1\columnwidth]{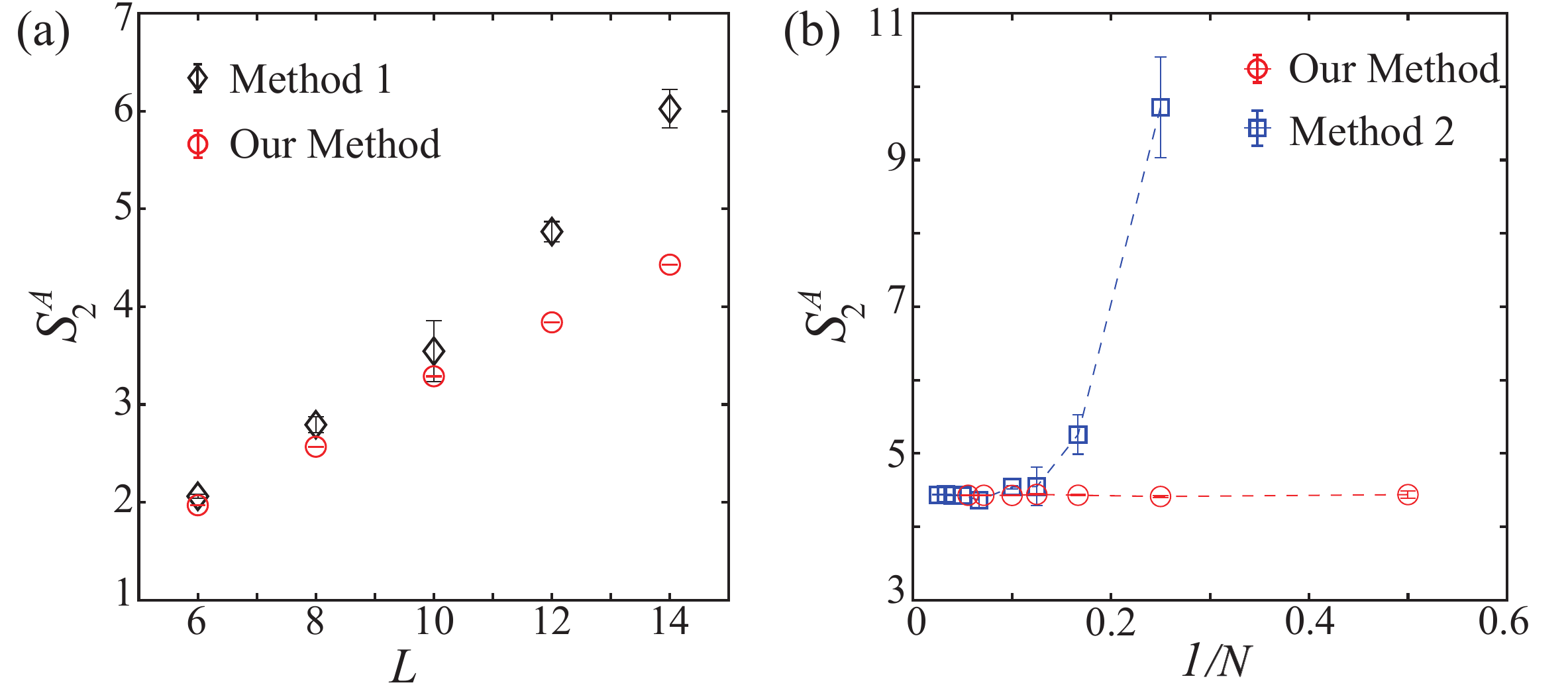}
\caption{(a) The $S_2^A$ obtained with method developed by Grover (Method 1)~\cite{groverEntanglement2013} and in this paper. (b) The $1/N$ extrapolation of $S_2^A$ for method developed by D’Emidio et al (Method 2)~\cite{demidioUniversal2022} and our method. It is important to note that the data obtained using Method 1 do not converge at larger $L$ due to inadequate sampling, which leads to unreliable error bars. Here $U=10$ and $L=14$ for $A_1$ entanglement region.}
\label{fig4}
\end{figure}

Below we compare the estimates of $S_2^A$ for $0$-flux Hubbard model at $U=10$ using the method developed by Grover (Method 1)~\cite{groverEntanglement2013}, the method developed by D’Emidio et al (Method 2)~\cite{demidioUniversal2022}, and our method within comparable CPU times.
As shown in Fig.~\ref{fig4}(a), the results obtained with Method 1 obviously deviate from the reliable values for $L>6$, and this deviation becomes more noticeable as the system sizes increase.
It is important to note that the errorbars of data obtained with Method 1 are not credible, because the mean values of Method 1 don't converge in principle. 
Furthermore, as shown in Fig.~\ref{fig4}(b), we plot the $1/N$ extrapolation of $S_2^A$ for $L=14$. 
As $N$ increases, the results obtained with Method 2 and our method gradually converge. 
However, Method 2 requires much larger $N$ to achieve the accuracy of our method.
More critically, we observe that the $S_2^A$ obtained with Method 2 exhibits worse accuracy than Method 1 when the total number of incremental processes is small. This is because the estimator in Method 2 cannot revert to Method 1 when the total number of incremental processes is reduced to 1, leading to bias. 
These limitations make Method 2 less convenient and reliable in practice. Therefore, our method should perform better in practical applications.
Further details comparing Method1, Method 2 and our method are discussed in the SM~\cite{suppl}.

{\it The direct method for calculating universal term of EE.---}
In (2+1)D SSB phase, when the boundary of a subsystem has sharp corners with angles $\theta_i$, one expects~\cite{metlitski2015entanglement,buenoUniversality2015,faulknerShape2016}
\begin{equation}
    S_2^A(L) = \mathcal{A} l + (s_G-b(\theta)) \ln{L} + c_A + \mathcal{O}(1/L),
\label{eq:corner}
\end{equation}
where $l$ is the boundary length of entanglement region $A$, $\mathcal{A}$ is a non-universal constant only related to the ground physics of system, $c_A$ is a non-universal constant related to the geometry of $A$, and the subleading logarithmic correction term is universal. The universal coefficient $s_G=N_g/2$ results from the $N_g$ Goldstone modes, and $b(\theta)$ consists of contributions from each corner: $b(\theta)= \sum_i b_A(\theta_i)$. 
Typically, $b(\theta)$ is much smaller than $s_G$, thus it is impossible to isolate $b$ from the fitting result of $s_G - b(\theta)$ due to the strong finite-size effects of $s_G$.
At a QCP, $s_G$ should equal 0.
While the complete form of the function $b(\theta)$ is not analytically known at the QCP, it is well-understood for free Dirac fermions. For instance, $b(\pi/2)=0.23936$ for spin-full free Dirac fermions on a square lattice with a square entanglement region $A_2$ and four $90^\circ$ corners, as depicted in Fig.\ref{fig1}(b)\cite{helmesUniversal2016}. However, the question of whether the universal coefficient $b$ remains consistent between the chiral Heisenberg Gross-Neveu QCP and free Dirac fermions remains open and presents a considerable challenge to address.

The challenge of extracting the universal coefficient $b$ from Eq.\eqref{eq:corner} is particularly prominent in interacting fermion systems, where it requires extremely high accuracy of $S_2^A$ and a large number of data points to perform function fitting. While the power incremental method developed in this letter can enhance the accuracy of $S_2^A$ to the requisite level, numerical simulations of interacting fermions are often restricted to small system sizes due to computational limitations. Consequently, only a few data points are available for function fitting, rendering the task of accurately fitting a complicated function such as Eq.\eqref{eq:corner} to obtain a reliable subleading universal term exceedingly challenging. This situation motivates us to develop a novel approach for straightforwardly calculating the universal term of the EE, circumventing the need for function fitting with limited data points.

It is well established that logarithmic corrections arising from corners vanish when the entanglement region has smooth boundaries~\cite{buenoUniversality2015,faulknerShape2016}, i.e., $b(\pi)=0$. As a result, if we select region $A$ as $A_3$ with a smooth boundary in SSB phase, as depicted in Fig.~\ref{fig1}(b), we have
$
S_2^{A_3}(L) = \mathcal{A} l_{A_3} + s_G \ln{L} + c_{A_3} + \mathcal{O}(1/L).
$
Additionally, we have
$
S_2^{A_2}(L) = \mathcal{A} l_{A_2} +(s_G- b(\pi/2)) \ln{L} + c_{A_2} + \mathcal{O}(1/L).
$
Considering that $l_{A_3} = l_{A_2}$, we can define the universal EE as
\begin{equation}\label{eq:Su}
S_u(L) \equiv S_2^{A_3}(L) - S_2^{A_2}(L) = b \ln {L} + c + \mathcal{O}(1/L),
\end{equation}
where $c$ is a non-universal constant.
Now, the universal coefficient $b$ becomes the leading term and is more easily extracted from fitting. 
We can straightforwardly calculate $S_u$ via the unbiased DQMC method in a single simulation as
\begin{equation}
e^{-S_u}=\frac{\sum_{{s_1},{s_2}} W_{s_1,s_2} \det g_{A_2}^{s_1, s_2} \frac{\det g_{A_3}^{s_1, s_2} }{\det g_{A_2}^{s_1, s_2} } }{\sum_{{s_1},{s_2}} W_{s_1,s_2} \det g_{A_2}^{s_1, s_2} }.
\end{equation}
Such a definition for the calculation of $S_u$ is also valid at the QCP and in the Dirac semi-metal (DSM) phase.

In the supplementary materials~\cite{suppl}, we further demonstrate that $S_u$ is also an exponential observable, akin to $S_2^A$, and we need to adapt our power incremental method to accurately estimate it. Notably, we have demonstrated that the computational cost for obtaining $S_u$ is lower than that for $S_2^{A_2}$. This implies that the straightforward technique for calculating the universal EE not only yields more reliable universal features but also is more cost-effective, thus constituting a significant advancement.
We would like to emphasize that the straightforward technique we propose here for calculating $S_u$ can be readily generalized to other methodologies for computing EE, like the incremental SWAP operator method and nonequilibrium method~\cite{Vincenzo2017,bulgarelliEntanglement2023,daliao2024extract}.

\begin{figure}[htp!]
\begin{minipage}{0.6\linewidth}
 \centerline{\includegraphics[width=1\linewidth]{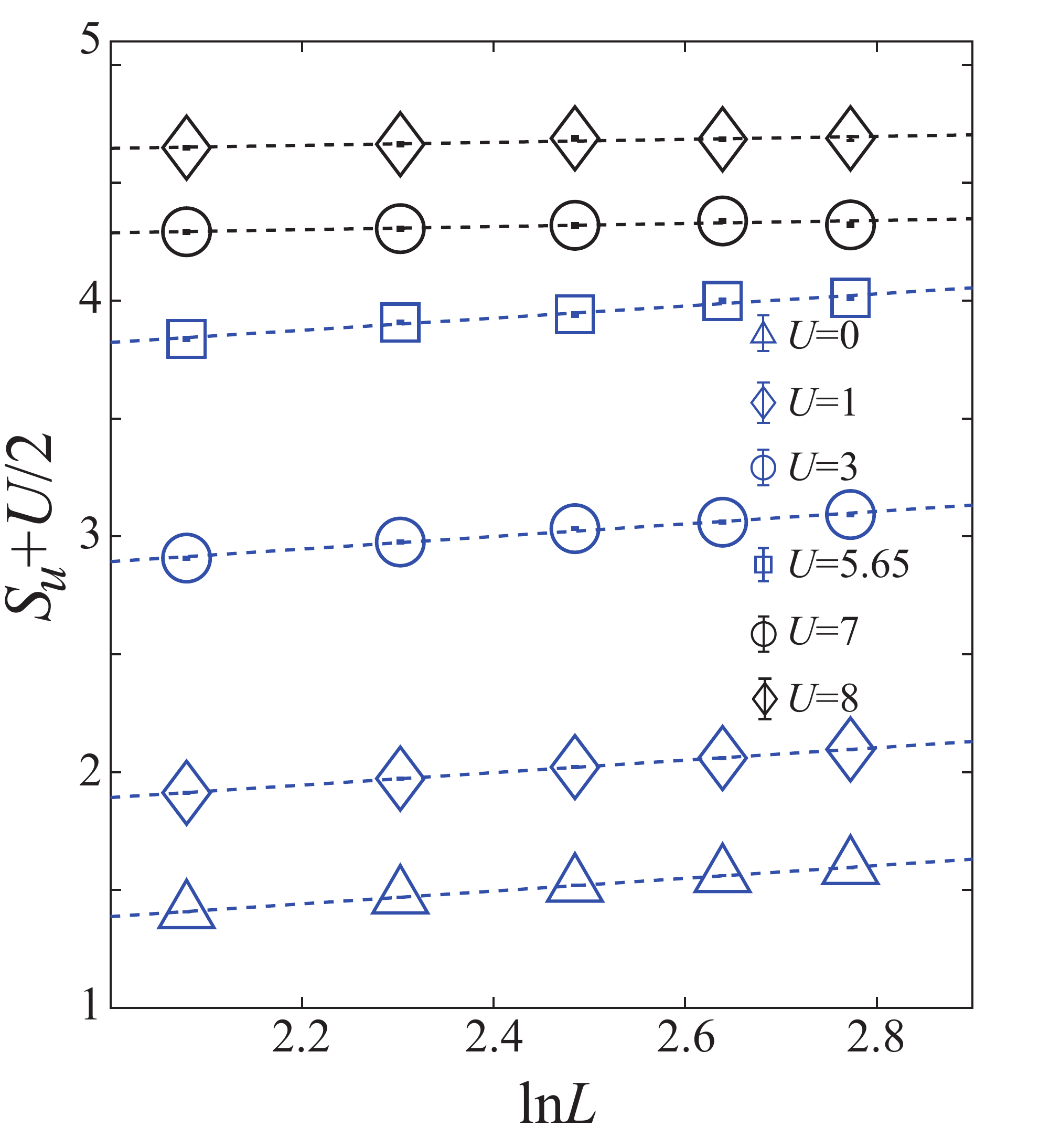}}
\end{minipage}
\hfill
\begin{minipage}{.38\linewidth}
	\begin{tabular}{lll}
			&\multicolumn{1}{c}{\ \ \ \ \ \ \ \ \ \ \ Table I} & \ \\
		\hline \hline &{$U$} & {$b(\pi/2)$} \\
        \hline & 0 &0.266(2)\\
               & 1 &0.264(3)\\
               & 3 &0.26(1)\\
               & 5.65 &0.26(2)\\
               & 7 &0.06(1)\\
               & 8 &0.06(1)\\
		\hline \hline
	\end{tabular}
\end{minipage}
\caption{
Left figure: $S_u$ of $\pi$-flux Hubbard model as function of $\ln L$ for different $U$, with a constant shift of $U/2$ added for clarity. 
Right Table I:
The universal constant $b(\pi/2)$ for different $U$.
We find consistency in the value of $b(\pi/2)$ for the Gross-Neveu QCP, the DSM phase with interactions and free Dirac fermions. While for AFM phase, $b(\pi/2)$ is comparable with previous prediction.
}
\label{fig5}
\end{figure}

{\it The universal term of EE for $\pi$-flux Hubbard model.---}
Next we move to the $\pi$-flux Hubbard model at half-filling. The Hamiltonian of $\pi$-flux Hubbard model is given by
\begin{equation}
H=-\sum_{\langle i j\rangle, \sigma} \left( t_{ij}  c_{i \sigma}^{\dagger} c_{j \sigma}+\text { H.c. }\right)+\frac{U}{2} \sum_{i}\left(n_{i\uparrow}+n_{i\downarrow}-1\right)^2,
\label{eq:piflux}
\end{equation}
where $t_{ij}=te^{i\phi_{ij}}$ represents the nearest-neighbor hopping amplitude, with $t=1$ set as the energy unit. 
When a single electron hops around each plaquette $\square$, it acquires a flux $\phi=\sum_{\square}\phi_{ij}$. Our focus is on the case $\phi=\pi$, as illustrated in Fig.~\ref{fig1} (b), which leads to the formation of Dirac cones in the first Brillouin zone.
This model has been extensively studied in the literature ~\cite{otsukaUniversal2016}. A quantum phase transition occurs from a DSM to an antiferromagnetic (AFM) phase at the QCP $U_c=5.65$, which belongs to the chiral Heisenberg Gross-Neveu universality class. 
It is worth noting that the universal term of EE for the $\pi$-flux Hubbard model with interactions ($U > 0$) has remained elusive due to the lack of a reliable computational algorithm. Our developments, by providing robust and precise methods for calculating universal term of EE in strongly correlated fermionic systems, can potentially shed light on this long-standing issue. 

We calculate $S_u$ for the $\pi$-flux Hubbard model at $U<U_c$ ($U=0$, $U=1$, and $U=3$), $U=U_c$, and $U>U_c$ ($U=7$ and $U=8$), and plot $S_u$ versus $\ln L$ with $L=8,10,12,14$ and $16$, as depicted in Fig.\ref{fig5}. It is evident that $S_u$ follows a linear relationship with $\ln L$, consistent with theoretical predictions. We further fit the coefficient according to Eq.\eqref{eq:Su}, with results listed in the Table I.
It is clear that the extracted universal coefficient $b(\pi/2)$ remains consistent for $U<U_c$ and $U=U_c$, leading us to hypothesize that the universal feature of the EE is preserved between free Dirac fermions, the DSM phase with interactions, and the Gross-Neveu QCP.
We notice that our fitted $b(\pi/2)=0.266(2)$ for $U=0$ is a little larger than the predicted value $0.23936$, which is due to the finite-size effect. We successfully extrapolate $b(\pi/2)$ to $0.23936$ for $U=0$ in the thermodynamic limit, the results are shown in the supplementary materials~\cite{suppl}~\cite{suppl}.

Moreover, we investigate the universal term of EE in the AFM-SSB phase. Two Goldstone bosons determine the exact value of the universal constant from EE in this ordered phase.
It is predicted that this universal constant of the $S^{A_2}_2$ originates from both Goldstone modes $s_G=1$ ~\cite{metlitski2015entanglement} and corner effects $b_{\text{Boson}}=0.052$~\cite{laflorencieSpin-wave2015,helmesUniversal2016}. 
The former is typically much larger than the latter and displays strong finite-size effects~\cite{dengImproved2023}, rendering the isolation of $b_{\text{Boson}}$ from $s_G$ unattainable by directly fitting $S^{A_2}_2$.
Our straightforward technique for calculating $S_u$ could cancel out the area-law term and the logarithmic contribution from Goldstone modes of $S^{A_2}_2$, exposing the universal corner contributions for the first time as the leading term, thus yielding unprecedented numerical accuracy for $b_{\text{Boson}}$ in the AFM phase of the $\pi$-flux Hubbard model.
Our extracted values of $b(\pi/2)=0.06(1)$ and $b(\pi/2)=0.06(1)$ at $U=7$ and $U=8$, respectively, are consistent with the theoretical prediction. These results further confirm the reliability of our method.

{\it Discussions.---}
We have devised a power incremental algorithm for calculating EE and we further propose a technique to straightforwardly estimate the universal term of EE.
Our algorithm is highly effective due to the reduction of the coefficient of variation during important sampling, which decreases from exponential growth to power-law decay.
The total number of incremental processes $N$ can be quantitatively determined, and each incremental process can be performed in parallel without bias, maintaining the computational time at the same order as the standard DQMC method.
These benefits make our algorithm more practical for research applications.
Moreover, we have revealed consistency between the universal terms to EE from sharp corners between free Dirac fermions, the DSM phase with interactions, and the Gross-Neveu QCP.
Additionally, our method can be extended to calculate other exponential observables, such as partition functions, entanglement spectra, and negativity in correlated boson/spin systems, as well as other general functionals of determinants of Green’s functions in interacting fermion systems.
Accurately measuring these elusive observables has the potential to provide inspiration for physical theory developments and guide future experimental directions.


\vskip4mm
\noindent
\textbf{ACKNOWLEDGEMENTS} 

\noindent
We thank Zi Yang Meng, Yang Qi, Zheng Yan for helpful discussions.
We acknowledges support from National Natural Science Foundation of China (Grant No. 12404282 and No. 12247114)
The author also acknowledges Beijng PARATERA Tech Co.,Ltd.~\cite{paratera} for providing HPC resources that have contributed to the research results reported in this paper.

\bibliography{ref.bib}

\begin{thebibliography}{35}%
\makeatletter
\providecommand \@ifxundefined [1]{%
 \@ifx{#1\undefined}
}%
\providecommand \@ifnum [1]{%
 \ifnum #1\expandafter \@firstoftwo
 \else \expandafter \@secondoftwo
 \fi
}%
\providecommand \@ifx [1]{%
 \ifx #1\expandafter \@firstoftwo
 \else \expandafter \@secondoftwo
 \fi
}%
\providecommand \natexlab [1]{#1}%
\providecommand \enquote  [1]{``#1''}%
\providecommand \bibnamefont  [1]{#1}%
\providecommand \bibfnamefont [1]{#1}%
\providecommand \citenamefont [1]{#1}%
\providecommand \href@noop [0]{\@secondoftwo}%
\providecommand \href [0]{\begingroup \@sanitize@url \@href}%
\providecommand \@href[1]{\@@startlink{#1}\@@href}%
\providecommand \@@href[1]{\endgroup#1\@@endlink}%
\providecommand \@sanitize@url [0]{\catcode `\\12\catcode `\$12\catcode
  `\&12\catcode `\#12\catcode `\^12\catcode `\_12\catcode `\%12\relax}%
\providecommand \@@startlink[1]{}%
\providecommand \@@endlink[0]{}%
\providecommand \url  [0]{\begingroup\@sanitize@url \@url }%
\providecommand \@url [1]{\endgroup\@href {#1}{\urlprefix }}%
\providecommand \urlprefix  [0]{URL }%
\providecommand \Eprint [0]{\href }%
\providecommand \doibase [0]{https://doi.org/}%
\providecommand \selectlanguage [0]{\@gobble}%
\providecommand \bibinfo  [0]{\@secondoftwo}%
\providecommand \bibfield  [0]{\@secondoftwo}%
\providecommand \translation [1]{[#1]}%
\providecommand \BibitemOpen [0]{}%
\providecommand \bibitemStop [0]{}%
\providecommand \bibitemNoStop [0]{.\EOS\space}%
\providecommand \EOS [0]{\spacefactor3000\relax}%
\providecommand \BibitemShut  [1]{\csname bibitem#1\endcsname}%
\let\auto@bib@innerbib\@empty
\bibitem [{\citenamefont {Klebanov}\ \emph {et~al.}(2012)\citenamefont
  {Klebanov}, \citenamefont {Pufu}, \citenamefont {Sachdev},\ and\
  \citenamefont {Safdi}}]{klebanovRenyi2012}%
  \BibitemOpen
  \bibfield  {author} {\bibinfo {author} {\bibfnamefont {I.~R.}\ \bibnamefont
  {Klebanov}}, \bibinfo {author} {\bibfnamefont {S.~S.}\ \bibnamefont {Pufu}},
  \bibinfo {author} {\bibfnamefont {S.}~\bibnamefont {Sachdev}},\ and\ \bibinfo
  {author} {\bibfnamefont {B.~R.}\ \bibnamefont {Safdi}},\ }\href
  {https://doi.org/10.1007/JHEP04(2012)074} {\bibfield  {journal} {\bibinfo
  {journal} {Journal of High Energy Physics}\ }\textbf {\bibinfo {volume}
  {2012}},\ \bibinfo {pages} {74} (\bibinfo {year} {2012})}\BibitemShut
  {NoStop}%
\bibitem [{\citenamefont {Altman}\ and\ \citenamefont
  {Vosk}(2015)}]{altmanUniversal2015}%
  \BibitemOpen
  \bibfield  {author} {\bibinfo {author} {\bibfnamefont {E.}~\bibnamefont
  {Altman}}\ and\ \bibinfo {author} {\bibfnamefont {R.}~\bibnamefont {Vosk}},\
  }\href {https://doi.org/10.1146/annurev-conmatphys-031214-014701} {\bibfield
  {journal} {\bibinfo  {journal} {Annual Review of Condensed Matter Physics}\
  }\textbf {\bibinfo {volume} {6}},\ \bibinfo {pages} {383} (\bibinfo {year}
  {2015})}\BibitemShut {NoStop}%
\bibitem [{\citenamefont {Eisert}\ \emph {et~al.}(2010)\citenamefont {Eisert},
  \citenamefont {Cramer},\ and\ \citenamefont {Plenio}}]{eisertColloquium2010}%
  \BibitemOpen
  \bibfield  {author} {\bibinfo {author} {\bibfnamefont {J.}~\bibnamefont
  {Eisert}}, \bibinfo {author} {\bibfnamefont {M.}~\bibnamefont {Cramer}},\
  and\ \bibinfo {author} {\bibfnamefont {M.~B.}\ \bibnamefont {Plenio}},\
  }\href {https://doi.org/10.1103/RevModPhys.82.277} {\bibfield  {journal}
  {\bibinfo  {journal} {Reviews of Modern Physics}\ }\textbf {\bibinfo {volume}
  {82}},\ \bibinfo {pages} {277} (\bibinfo {year} {2010})}\BibitemShut
  {NoStop}%
\bibitem [{\citenamefont {Casini}\ and\ \citenamefont
  {Huerta}(2007)}]{casiniUniversal2007}%
  \BibitemOpen
  \bibfield  {author} {\bibinfo {author} {\bibfnamefont {H.}~\bibnamefont
  {Casini}}\ and\ \bibinfo {author} {\bibfnamefont {M.}~\bibnamefont
  {Huerta}},\ }\href {https://doi.org/10.1016/j.nuclphysb.2006.12.012}
  {\bibfield  {journal} {\bibinfo  {journal} {Nuclear Physics B}\ }\textbf
  {\bibinfo {volume} {764}},\ \bibinfo {pages} {183} (\bibinfo {year}
  {2007})}\BibitemShut {NoStop}%
\bibitem [{\citenamefont {Vidal}\ \emph {et~al.}(2003)\citenamefont {Vidal},
  \citenamefont {Latorre}, \citenamefont {Rico},\ and\ \citenamefont
  {Kitaev}}]{vidalEntanglement2003}%
  \BibitemOpen
  \bibfield  {author} {\bibinfo {author} {\bibfnamefont {G.}~\bibnamefont
  {Vidal}}, \bibinfo {author} {\bibfnamefont {J.~I.}\ \bibnamefont {Latorre}},
  \bibinfo {author} {\bibfnamefont {E.}~\bibnamefont {Rico}},\ and\ \bibinfo
  {author} {\bibfnamefont {A.}~\bibnamefont {Kitaev}},\ }\href
  {https://doi.org/10.1103/PhysRevLett.90.227902} {\bibfield  {journal}
  {\bibinfo  {journal} {Physical Review Letters}\ }\textbf {\bibinfo {volume}
  {90}},\ \bibinfo {pages} {227902} (\bibinfo {year} {2003})}\BibitemShut
  {NoStop}%
\bibitem [{\citenamefont {Calabrese}\ and\ \citenamefont
  {Cardy}(2004)}]{calabreseEntanglement2004}%
  \BibitemOpen
  \bibfield  {author} {\bibinfo {author} {\bibfnamefont {P.}~\bibnamefont
  {Calabrese}}\ and\ \bibinfo {author} {\bibfnamefont {J.}~\bibnamefont
  {Cardy}},\ }\href {https://doi.org/10.1088/1742-5468/2004/06/P06002}
  {\bibfield  {journal} {\bibinfo  {journal} {Journal of Statistical Mechanics:
  Theory and Experiment}\ }\textbf {\bibinfo {volume} {2004}},\ \bibinfo
  {pages} {P06002} (\bibinfo {year} {2004})}\BibitemShut {NoStop}%
\bibitem [{\citenamefont {Korepin}(2004)}]{korepinUniversality2004a}%
  \BibitemOpen
  \bibfield  {author} {\bibinfo {author} {\bibfnamefont {V.~E.}\ \bibnamefont
  {Korepin}},\ }\href {https://doi.org/10.1103/PhysRevLett.92.096402}
  {\bibfield  {journal} {\bibinfo  {journal} {Physical Review Letters}\
  }\textbf {\bibinfo {volume} {92}},\ \bibinfo {pages} {096402} (\bibinfo
  {year} {2004})}\BibitemShut {NoStop}%
\bibitem [{\citenamefont {Song}\ \emph {et~al.}(2011)\citenamefont {Song},
  \citenamefont {Laflorencie}, \citenamefont {Rachel},\ and\ \citenamefont
  {Le~Hur}}]{songEntanglement2011}%
  \BibitemOpen
  \bibfield  {author} {\bibinfo {author} {\bibfnamefont {H.~F.}\ \bibnamefont
  {Song}}, \bibinfo {author} {\bibfnamefont {N.}~\bibnamefont {Laflorencie}},
  \bibinfo {author} {\bibfnamefont {S.}~\bibnamefont {Rachel}},\ and\ \bibinfo
  {author} {\bibfnamefont {K.}~\bibnamefont {Le~Hur}},\ }\href
  {https://doi.org/10.1103/PhysRevB.83.224410} {\bibfield  {journal} {\bibinfo
  {journal} {Physical Review B}\ }\textbf {\bibinfo {volume} {83}},\ \bibinfo
  {pages} {224410} (\bibinfo {year} {2011})}\BibitemShut {NoStop}%
\bibitem [{\citenamefont {Helmes}\ \emph {et~al.}(2016)\citenamefont {Helmes},
  \citenamefont {Hayward~Sierens}, \citenamefont {Chandran}, \citenamefont
  {{Witczak-Krempa}},\ and\ \citenamefont {Melko}}]{helmesUniversal2016}%
  \BibitemOpen
  \bibfield  {author} {\bibinfo {author} {\bibfnamefont {J.}~\bibnamefont
  {Helmes}}, \bibinfo {author} {\bibfnamefont {L.~E.}\ \bibnamefont
  {Hayward~Sierens}}, \bibinfo {author} {\bibfnamefont {A.}~\bibnamefont
  {Chandran}}, \bibinfo {author} {\bibfnamefont {W.}~\bibnamefont
  {{Witczak-Krempa}}},\ and\ \bibinfo {author} {\bibfnamefont {R.~G.}\
  \bibnamefont {Melko}},\ }\href {https://doi.org/10.1103/PhysRevB.94.125142}
  {\bibfield  {journal} {\bibinfo  {journal} {Physical Review B}\ }\textbf
  {\bibinfo {volume} {94}},\ \bibinfo {pages} {125142} (\bibinfo {year}
  {2016})}\BibitemShut {NoStop}%
\bibitem [{\citenamefont {D'Emidio}\ \emph {et~al.}(2024)\citenamefont
  {D'Emidio}, \citenamefont {Or\'us}, \citenamefont {Laflorencie},\ and\
  \citenamefont {de~Juan}}]{demidioUniversal2022}%
  \BibitemOpen
  \bibfield  {author} {\bibinfo {author} {\bibfnamefont {J.}~\bibnamefont
  {D'Emidio}}, \bibinfo {author} {\bibfnamefont {R.}~\bibnamefont {Or\'us}},
  \bibinfo {author} {\bibfnamefont {N.}~\bibnamefont {Laflorencie}},\ and\
  \bibinfo {author} {\bibfnamefont {F.}~\bibnamefont {de~Juan}},\ }\href
  {https://doi.org/10.1103/PhysRevLett.132.076502} {\bibfield  {journal}
  {\bibinfo  {journal} {Phys. Rev. Lett.}\ }\textbf {\bibinfo {volume} {132}},\
  \bibinfo {pages} {076502} (\bibinfo {year} {2024})}\BibitemShut {NoStop}%
\bibitem [{\citenamefont {Da~Liao}\ \emph {et~al.}(2023)\citenamefont
  {Da~Liao}, \citenamefont {Pan}, \citenamefont {Jiang}, \citenamefont {Qi},\
  and\ \citenamefont {Meng}}]{daliaoteaching2023}%
  \BibitemOpen
  \bibfield  {author} {\bibinfo {author} {\bibfnamefont {Y.}~\bibnamefont
  {Da~Liao}}, \bibinfo {author} {\bibfnamefont {G.}~\bibnamefont {Pan}},
  \bibinfo {author} {\bibfnamefont {W.}~\bibnamefont {Jiang}}, \bibinfo
  {author} {\bibfnamefont {Y.}~\bibnamefont {Qi}},\ and\ \bibinfo {author}
  {\bibfnamefont {Z.~Y.}\ \bibnamefont {Meng}},\ }\href
  {http://arxiv.org/abs/2302.11742} {\bibinfo {title} {The teaching from
  entanglement: {{2D}} deconfined quantum critical points are not conformal}}
  (\bibinfo {year} {2023}),\ \Eprint {https://arxiv.org/abs/2302.11742}
  {arxiv:2302.11742 [cond-mat, physics:math-ph, physics:physics,
  physics:quant-ph]} \BibitemShut {NoStop}%
\bibitem [{\citenamefont {Metlitski}\ and\ \citenamefont
  {Grover}(2015)}]{metlitski2015entanglement}%
  \BibitemOpen
  \bibfield  {author} {\bibinfo {author} {\bibfnamefont {M.~A.}\ \bibnamefont
  {Metlitski}}\ and\ \bibinfo {author} {\bibfnamefont {T.}~\bibnamefont
  {Grover}},\ }\href {https://doi.org/10.48550/arXiv.1112.5166} {\bibinfo
  {title} {Entanglement entropy of systems with spontaneously broken continuous
  symmetry}} (\bibinfo {year} {2015}),\ \Eprint
  {https://arxiv.org/abs/1112.5166} {arXiv:1112.5166 [cond-mat.str-el]}
  \BibitemShut {NoStop}%
\bibitem [{\citenamefont {Kallin}\ \emph {et~al.}(2011)\citenamefont {Kallin},
  \citenamefont {Hastings}, \citenamefont {Melko},\ and\ \citenamefont
  {Singh}}]{kallinAnomalies2011}%
  \BibitemOpen
  \bibfield  {author} {\bibinfo {author} {\bibfnamefont {A.~B.}\ \bibnamefont
  {Kallin}}, \bibinfo {author} {\bibfnamefont {M.~B.}\ \bibnamefont
  {Hastings}}, \bibinfo {author} {\bibfnamefont {R.~G.}\ \bibnamefont
  {Melko}},\ and\ \bibinfo {author} {\bibfnamefont {R.~R.~P.}\ \bibnamefont
  {Singh}},\ }\href {https://doi.org/10.1103/PhysRevB.84.165134} {\bibfield
  {journal} {\bibinfo  {journal} {Physical Review B}\ }\textbf {\bibinfo
  {volume} {84}},\ \bibinfo {pages} {165134} (\bibinfo {year}
  {2011})}\BibitemShut {NoStop}%
\bibitem [{\citenamefont {D'Emidio}(2020)}]{emidioEntanglement2020}%
  \BibitemOpen
  \bibfield  {author} {\bibinfo {author} {\bibfnamefont {J.}~\bibnamefont
  {D'Emidio}},\ }\href {https://doi.org/10.1103/PhysRevLett.124.110602}
  {\bibfield  {journal} {\bibinfo  {journal} {Physical Review Letters}\
  }\textbf {\bibinfo {volume} {124}},\ \bibinfo {pages} {110602} (\bibinfo
  {year} {2020})}\BibitemShut {NoStop}%
\bibitem [{\citenamefont {Zhao}\ \emph {et~al.}(2022)\citenamefont {Zhao},
  \citenamefont {Chen}, \citenamefont {Wang}, \citenamefont {Yan},
  \citenamefont {Cheng},\ and\ \citenamefont {Meng}}]{zhaoMeasuring2022}%
  \BibitemOpen
  \bibfield  {author} {\bibinfo {author} {\bibfnamefont {J.}~\bibnamefont
  {Zhao}}, \bibinfo {author} {\bibfnamefont {B.-B.}\ \bibnamefont {Chen}},
  \bibinfo {author} {\bibfnamefont {Y.-C.}\ \bibnamefont {Wang}}, \bibinfo
  {author} {\bibfnamefont {Z.}~\bibnamefont {Yan}}, \bibinfo {author}
  {\bibfnamefont {M.}~\bibnamefont {Cheng}},\ and\ \bibinfo {author}
  {\bibfnamefont {Z.~Y.}\ \bibnamefont {Meng}},\ }\href
  {https://doi.org/10.1038/s41535-022-00476-0} {\bibfield  {journal} {\bibinfo
  {journal} {npj Quantum Materials}\ }\textbf {\bibinfo {volume} {7}},\
  \bibinfo {pages} {69} (\bibinfo {year} {2022})}\BibitemShut {NoStop}%
\bibitem [{\citenamefont {Pan}\ \emph {et~al.}(2023)\citenamefont {Pan},
  \citenamefont {Da~Liao}, \citenamefont {Jiang}, \citenamefont {D'Emidio},
  \citenamefont {Qi},\ and\ \citenamefont {Meng}}]{panComputing2023}%
  \BibitemOpen
  \bibfield  {author} {\bibinfo {author} {\bibfnamefont {G.}~\bibnamefont
  {Pan}}, \bibinfo {author} {\bibfnamefont {Y.}~\bibnamefont {Da~Liao}},
  \bibinfo {author} {\bibfnamefont {W.}~\bibnamefont {Jiang}}, \bibinfo
  {author} {\bibfnamefont {J.}~\bibnamefont {D'Emidio}}, \bibinfo {author}
  {\bibfnamefont {Y.}~\bibnamefont {Qi}},\ and\ \bibinfo {author}
  {\bibfnamefont {Z.~Y.}\ \bibnamefont {Meng}},\ }\href
  {https://doi.org/10.1103/PhysRevB.108.L081123} {\bibfield  {journal}
  {\bibinfo  {journal} {Phys. Rev. B}\ }\textbf {\bibinfo {volume} {108}},\
  \bibinfo {pages} {L081123} (\bibinfo {year} {2023})}\BibitemShut {NoStop}%
\bibitem [{\citenamefont {Deng}\ \emph {et~al.}(2023)\citenamefont {Deng},
  \citenamefont {Liu}, \citenamefont {Guo},\ and\ \citenamefont
  {Lin}}]{dengImproved2023}%
  \BibitemOpen
  \bibfield  {author} {\bibinfo {author} {\bibfnamefont {Z.}~\bibnamefont
  {Deng}}, \bibinfo {author} {\bibfnamefont {L.}~\bibnamefont {Liu}}, \bibinfo
  {author} {\bibfnamefont {W.}~\bibnamefont {Guo}},\ and\ \bibinfo {author}
  {\bibfnamefont {H.~Q.}\ \bibnamefont {Lin}},\ }\href
  {https://doi.org/10.1103/PhysRevB.108.125144} {\bibfield  {journal} {\bibinfo
   {journal} {Phys. Rev. B}\ }\textbf {\bibinfo {volume} {108}},\ \bibinfo
  {pages} {125144} (\bibinfo {year} {2023})}\BibitemShut {NoStop}%
\bibitem [{\citenamefont {Grover}(2013)}]{groverEntanglement2013}%
  \BibitemOpen
  \bibfield  {author} {\bibinfo {author} {\bibfnamefont {T.}~\bibnamefont
  {Grover}},\ }\href {https://doi.org/10.1103/PhysRevLett.111.130402}
  {\bibfield  {journal} {\bibinfo  {journal} {Physical Review Letters}\
  }\textbf {\bibinfo {volume} {111}},\ \bibinfo {pages} {130402} (\bibinfo
  {year} {2013})}\BibitemShut {NoStop}%
\bibitem [{\citenamefont {Broecker}\ and\ \citenamefont
  {Trebst}(2014)}]{broeckerRenyi2014}%
  \BibitemOpen
  \bibfield  {author} {\bibinfo {author} {\bibfnamefont {P.}~\bibnamefont
  {Broecker}}\ and\ \bibinfo {author} {\bibfnamefont {S.}~\bibnamefont
  {Trebst}},\ }\href {https://doi.org/10.1088/1742-5468/2014/08/P08015}
  {\bibfield  {journal} {\bibinfo  {journal} {Journal of Statistical Mechanics:
  Theory and Experiment}\ }\textbf {\bibinfo {volume} {2014}},\ \bibinfo
  {pages} {P08015} (\bibinfo {year} {2014})}\BibitemShut {NoStop}%
\bibitem [{\citenamefont {Wang}\ and\ \citenamefont
  {Troyer}(2014)}]{wangRenyi2014}%
  \BibitemOpen
  \bibfield  {author} {\bibinfo {author} {\bibfnamefont {L.}~\bibnamefont
  {Wang}}\ and\ \bibinfo {author} {\bibfnamefont {M.}~\bibnamefont {Troyer}},\
  }\href {https://doi.org/10.1103/PhysRevLett.113.110401} {\bibfield  {journal}
  {\bibinfo  {journal} {Physical Review Letters}\ }\textbf {\bibinfo {volume}
  {113}},\ \bibinfo {pages} {110401} (\bibinfo {year} {2014})}\BibitemShut
  {NoStop}%
\bibitem [{\citenamefont {Qin}\ \emph {et~al.}(2022)\citenamefont {Qin},
  \citenamefont {Sch\"{a}fer}, \citenamefont {Andergassen}, \citenamefont
  {Corboz},\ and\ \citenamefont {Gull}}]{qinHubbard2022}%
  \BibitemOpen
  \bibfield  {author} {\bibinfo {author} {\bibfnamefont {M.}~\bibnamefont
  {Qin}}, \bibinfo {author} {\bibfnamefont {T.}~\bibnamefont {Sch\"{a}fer}},
  \bibinfo {author} {\bibfnamefont {S.}~\bibnamefont {Andergassen}}, \bibinfo
  {author} {\bibfnamefont {P.}~\bibnamefont {Corboz}},\ and\ \bibinfo {author}
  {\bibfnamefont {E.}~\bibnamefont {Gull}},\ }\href
  {https://doi.org/10.1146/annurev-conmatphys-090921-033948} {\bibfield
  {journal} {\bibinfo  {journal} {Annual Review of Condensed Matter Physics}\
  }\textbf {\bibinfo {volume} {13}},\ \bibinfo {pages} {275} (\bibinfo {year}
  {2022})},\ \Eprint
  {https://arxiv.org/abs/https://doi.org/10.1146/annurev-conmatphys-090921-033948}
  {https://doi.org/10.1146/annurev-conmatphys-090921-033948} \BibitemShut
  {NoStop}%
\bibitem [{\citenamefont {de~Forcrand}\ \emph {et~al.}(2001)\citenamefont
  {de~Forcrand}, \citenamefont {D'Elia},\ and\ \citenamefont
  {Pepe}}]{Forcrand2001}%
  \BibitemOpen
  \bibfield  {author} {\bibinfo {author} {\bibfnamefont {P.}~\bibnamefont
  {de~Forcrand}}, \bibinfo {author} {\bibfnamefont {M.}~\bibnamefont
  {D'Elia}},\ and\ \bibinfo {author} {\bibfnamefont {M.}~\bibnamefont {Pepe}},\
  }\href {https://doi.org/10.1103/PhysRevLett.86.1438} {\bibfield  {journal}
  {\bibinfo  {journal} {Phys. Rev. Lett.}\ }\textbf {\bibinfo {volume} {86}},\
  \bibinfo {pages} {1438} (\bibinfo {year} {2001})}\BibitemShut {NoStop}%
\bibitem [{sup()}]{suppl}%
  \BibitemOpen
  \href@noop {} {\bibinfo  {journal} {In this Supplementary Material, we
  provide an introduction to determinant quantum Monte Carlo (DQMC) method and
  the replica trick for the entanglement entropy (EE), a more detailed
  description of the direct method for calculating the universal term of the
  EE, denoted as $S_u$, and further demonstrate that $S_u$ is an exponential
  observable. We also discuss the finite-size effect of $S_u$ at $U=0$ and the
  positivity of Grover matrix determinant.}\ }\BibitemShut {NoStop}%
\bibitem [{\citenamefont {Bueno}\ \emph {et~al.}(2015)\citenamefont {Bueno},
  \citenamefont {Myers},\ and\ \citenamefont
  {{Witczak-Krempa}}}]{buenoUniversality2015}%
  \BibitemOpen
\bibfield  {journal} {  }\bibfield  {author} {\bibinfo {author} {\bibfnamefont
  {P.}~\bibnamefont {Bueno}}, \bibinfo {author} {\bibfnamefont {R.~C.}\
  \bibnamefont {Myers}},\ and\ \bibinfo {author} {\bibfnamefont
  {W.}~\bibnamefont {{Witczak-Krempa}}},\ }\href
  {https://doi.org/10.1103/PhysRevLett.115.021602} {\bibfield  {journal}
  {\bibinfo  {journal} {Physical Review Letters}\ }\textbf {\bibinfo {volume}
  {115}},\ \bibinfo {pages} {021602} (\bibinfo {year} {2015})}\BibitemShut
  {NoStop}%
\bibitem [{\citenamefont {Faulkner}\ \emph {et~al.}(2016)\citenamefont
  {Faulkner}, \citenamefont {Leigh},\ and\ \citenamefont
  {Parrikar}}]{faulknerShape2016}%
  \BibitemOpen
  \bibfield  {author} {\bibinfo {author} {\bibfnamefont {T.}~\bibnamefont
  {Faulkner}}, \bibinfo {author} {\bibfnamefont {R.~G.}\ \bibnamefont
  {Leigh}},\ and\ \bibinfo {author} {\bibfnamefont {O.}~\bibnamefont
  {Parrikar}},\ }\href {https://doi.org/10.1007/JHEP04(2016)088} {\bibfield
  {journal} {\bibinfo  {journal} {Journal of High Energy Physics}\ }\textbf
  {\bibinfo {volume} {2016}},\ \bibinfo {pages} {1} (\bibinfo {year}
  {2016})}\BibitemShut {NoStop}%
\bibitem [{\citenamefont {Alba}(2017)}]{Vincenzo2017}%
  \BibitemOpen
  \bibfield  {author} {\bibinfo {author} {\bibfnamefont {V.}~\bibnamefont
  {Alba}},\ }\href {https://doi.org/10.1103/PhysRevE.95.062132} {\bibfield
  {journal} {\bibinfo  {journal} {Phys. Rev. E}\ }\textbf {\bibinfo {volume}
  {95}},\ \bibinfo {pages} {062132} (\bibinfo {year} {2017})}\BibitemShut
  {NoStop}%
\bibitem [{\citenamefont {Bulgarelli}\ and\ \citenamefont
  {Panero}(2023)}]{bulgarelliEntanglement2023}%
  \BibitemOpen
  \bibfield  {author} {\bibinfo {author} {\bibfnamefont {A.}~\bibnamefont
  {Bulgarelli}}\ and\ \bibinfo {author} {\bibfnamefont {M.}~\bibnamefont
  {Panero}},\ }\href {https://doi.org/10.1007/JHEP06(2023)030} {\bibfield
  {journal} {\bibinfo  {journal} {Journal of High Energy Physics}\ }\textbf
  {\bibinfo {volume} {2023}},\ \bibinfo {pages} {30} (\bibinfo {year}
  {2023})}\BibitemShut {NoStop}%
\bibitem [{\citenamefont {Da~Liao}\ \emph {et~al.}(2024)\citenamefont
  {Da~Liao}, \citenamefont {Song}, \citenamefont {Zhao},\ and\ \citenamefont
  {Meng}}]{daliao2024extract}%
  \BibitemOpen
  \bibfield  {author} {\bibinfo {author} {\bibfnamefont {Y.}~\bibnamefont
  {Da~Liao}}, \bibinfo {author} {\bibfnamefont {M.}~\bibnamefont {Song}},
  \bibinfo {author} {\bibfnamefont {J.}~\bibnamefont {Zhao}},\ and\ \bibinfo
  {author} {\bibfnamefont {Z.~Y.}\ \bibnamefont {Meng}},\ }\href
  {https://doi.org/10.1103/PhysRevB.110.235111} {\bibfield  {journal} {\bibinfo
   {journal} {Phys. Rev. B}\ }\textbf {\bibinfo {volume} {110}},\ \bibinfo
  {pages} {235111} (\bibinfo {year} {2024})}\BibitemShut {NoStop}%
\bibitem [{\citenamefont {Otsuka}\ \emph {et~al.}(2016)\citenamefont {Otsuka},
  \citenamefont {Yunoki},\ and\ \citenamefont {Sorella}}]{otsukaUniversal2016}%
  \BibitemOpen
  \bibfield  {author} {\bibinfo {author} {\bibfnamefont {Y.}~\bibnamefont
  {Otsuka}}, \bibinfo {author} {\bibfnamefont {S.}~\bibnamefont {Yunoki}},\
  and\ \bibinfo {author} {\bibfnamefont {S.}~\bibnamefont {Sorella}},\ }\href
  {https://doi.org/10.1103/PhysRevX.6.011029} {\bibfield  {journal} {\bibinfo
  {journal} {Physical Review X}\ }\textbf {\bibinfo {volume} {6}},\ \bibinfo
  {pages} {011029} (\bibinfo {year} {2016})}\BibitemShut {NoStop}%
\bibitem [{\citenamefont {Laflorencie}\ \emph {et~al.}(2015)\citenamefont
  {Laflorencie}, \citenamefont {Luitz},\ and\ \citenamefont
  {Alet}}]{laflorencieSpin-wave2015}%
  \BibitemOpen
  \bibfield  {author} {\bibinfo {author} {\bibfnamefont {N.}~\bibnamefont
  {Laflorencie}}, \bibinfo {author} {\bibfnamefont {D.~J.}\ \bibnamefont
  {Luitz}},\ and\ \bibinfo {author} {\bibfnamefont {F.}~\bibnamefont {Alet}},\
  }\href {https://doi.org/10.1103/PhysRevB.92.115126} {\bibfield  {journal}
  {\bibinfo  {journal} {Physical Review B}\ }\textbf {\bibinfo {volume} {92}},\
  \bibinfo {pages} {115126} (\bibinfo {year} {2015})}\BibitemShut {NoStop}%
\bibitem [{par()}]{paratera}%
  \BibitemOpen
  \href {https://www.paratera.com} {\bibinfo  {journal}
  {https://www.paratera.com}\ }\BibitemShut {NoStop}%
\bibitem [{\citenamefont {Assaad}\ and\ \citenamefont
  {Evertz}(2008)}]{assaadWorld-line2008}%
  \BibitemOpen
\bibfield  {journal} {  }\bibfield  {author} {\bibinfo {author} {\bibfnamefont
  {F.}~\bibnamefont {Assaad}}\ and\ \bibinfo {author} {\bibfnamefont
  {H.}~\bibnamefont {Evertz}},\ }in\ \href
  {https://doi.org/10.1007/978-3-540-74686-7_10} {\emph {\bibinfo {booktitle}
  {Computational {{Many-Particle Physics}}}}},\ Vol.\ \bibinfo {volume} {739},\
  \bibinfo {editor} {edited by\ \bibinfo {editor} {\bibfnamefont
  {H.}~\bibnamefont {Fehske}}, \bibinfo {editor} {\bibfnamefont
  {R.}~\bibnamefont {Schneider}},\ and\ \bibinfo {editor} {\bibfnamefont
  {A.}~\bibnamefont {Wei{\ss}e}}}\ (\bibinfo  {publisher} {Springer Berlin
  Heidelberg},\ \bibinfo {address} {Berlin, Heidelberg},\ \bibinfo {year}
  {2008})\ pp.\ \bibinfo {pages} {277--356}\BibitemShut {NoStop}%
\bibitem [{\citenamefont {{Yuan Da Liao}}\ \emph {et~al.}(2021)\citenamefont
  {{Yuan Da Liao}}, \citenamefont {{Jian Kang}}, \citenamefont {{Clara N.
  Brei{\o}}}, \citenamefont {{Xiao Yan Xu}}, \citenamefont {{Han-Qing Wu}},
  \citenamefont {{Brian M. Andersen}}, \citenamefont {{Rafael M. Fernandes}},\
  and\ \citenamefont {{Zi Yang Meng}}}]{yuandaliaoCorrelationInduced2021}%
  \BibitemOpen
  \bibfield  {author} {\bibinfo {author} {\bibnamefont {{Yuan Da Liao}}},
  \bibinfo {author} {\bibnamefont {{Jian Kang}}}, \bibinfo {author}
  {\bibnamefont {{Clara N. Brei{\o}}}}, \bibinfo {author} {\bibnamefont {{Xiao
  Yan Xu}}}, \bibinfo {author} {\bibnamefont {{Han-Qing Wu}}}, \bibinfo
  {author} {\bibnamefont {{Brian M. Andersen}}}, \bibinfo {author}
  {\bibnamefont {{Rafael M. Fernandes}}},\ and\ \bibinfo {author} {\bibnamefont
  {{Zi Yang Meng}}},\ }\href {https://doi.org/10.1103/PhysRevX.11.011014}
  {\bibfield  {journal} {\bibinfo  {journal} {Physical Review X}\ }\textbf
  {\bibinfo {volume} {11}},\ \bibinfo {pages} {011014} (\bibinfo {year}
  {2021})}\BibitemShut {NoStop}%
\bibitem [{\citenamefont {Xu}\ and\ \citenamefont
  {Grover}(2021)}]{xuCompeting2021}%
  \BibitemOpen
  \bibfield  {author} {\bibinfo {author} {\bibfnamefont {X.~Y.}\ \bibnamefont
  {Xu}}\ and\ \bibinfo {author} {\bibfnamefont {T.}~\bibnamefont {Grover}},\
  }\href {https://doi.org/10.1103/PhysRevLett.126.217002} {\bibfield  {journal}
  {\bibinfo  {journal} {Phys. Rev. Lett.}\ }\textbf {\bibinfo {volume} {126}},\
  \bibinfo {pages} {217002} (\bibinfo {year} {2021})}\BibitemShut {NoStop}%
\bibitem [{\citenamefont {Wang}\ and\ \citenamefont
  {Xu}(2023)}]{wangEntanglement2023a}%
  \BibitemOpen
  \bibfield  {author} {\bibinfo {author} {\bibfnamefont {F.-H.}\ \bibnamefont
  {Wang}}\ and\ \bibinfo {author} {\bibfnamefont {X.~Y.}\ \bibnamefont {Xu}},\
  }\href {http://arxiv.org/abs/2312.14155} {\bibinfo {title} {Entanglement
  {{R}}{\textbackslash}'\{e\}nyi {{Negativity}} of {{Interacting Fermions}}
  from {{Quantum Monte Carlo Simulations}}}} (\bibinfo {year} {2023}),\ \Eprint
  {https://arxiv.org/abs/2312.14155} {arxiv:2312.14155 [cond-mat,
  physics:hep-lat, physics:quant-ph]} \BibitemShut {NoStop}%
\end{thebibliography}%
\bibliographystyle{apsrev4-2}

\clearpage
\onecolumngrid

\appendix
\setcounter{equation}{0}
\setcounter{figure}{0}
\setcounter{table}{0}
\setcounter{page}{1}
\makeatletter
\renewcommand{\theequation}{S\arabic{equation}}
\renewcommand{\thefigure}{S\arabic{figure}}
\renewcommand{\bibnumfmt}[1]{[S#1]}
\renewcommand{\citenumfont}[1]{S#1}
\setcounter{secnumdepth}{3}

\begin{center}
\textbf{Supplemental Material for ``Universal term of Entanglement Entropy in the $\pi$-flux Hubbard model''}
\end{center} 

\vskip4mm

In this Supplementary Material, we provide an introduction to determinant quantum Monte Carlo (DQMC)  method and the replica trick for the entanglement entropy (EE), a more detailed description of the direct method for calculating the universal term of the EE, denoted as $S_u$, and further demonstrate that $S_u$ is an exponential observable. We also discuss the finite-size effect of $S_u$ at $U=0$ and the positivity of Grover matrix determinant.

\section{Projector DQMC method and replica trick for the EE}
The second Rényi EE is a physical quantity that is well-defined at zero temperature. In this work, we use the projector DQMC method to investigate the ground state properties of the $0$- and $\pi$-flux Hubbard model with on-site interaction.
In projector DQMC, the Hamiltonian, as represented in Eq. (8) of the main text, can be expressed as $ H = H_T + H_U $, where the noninteracting part is $ H_T = -\sum_{\langle i j \rangle, \sigma} \left( t_{ij} c_{i \sigma}^{\dagger} c_{j \sigma} + \text{H.c.} \right) $ and the interacting part is $ H_U = \frac{U}{2} \sum_{i} \left( n_{i \uparrow} + n_{i \downarrow} - 1 \right)^2 $.
To obtain the ground state wave function $\vert \Psi_0 \rangle$, one can project a trial wave function $\vert \Psi_T \rangle$ with a projector length $\beta$ along the imaginary axis, given by $\vert \Psi_0 \rangle = \lim_{\beta \to \infty} e^{-\frac{\beta}{2} H} \vert \Psi_T \rangle$. The observables can then be calculated as follows:
\begin{equation}
\label{eq:observablepqmc}
\langle \hat{O} \rangle = \frac{\langle \Psi_0 \vert \hat{O} \vert \Psi_0 \rangle}{\langle \Psi_0 \vert \Psi_0 \rangle} 
						= \lim\limits_{\Theta \to \infty} \frac{\langle \Psi_T \vert  e^{-\frac{\beta}{2} {H}} \hat{O}  e^{-\frac{\beta}{2} {H}} \vert \Psi_T \rangle}{\langle \Psi_T \vert  e^{-\beta {H}} \vert \Psi_T \rangle} .
\end{equation}

Since $ H $ consists of non-interacting and interacting parts, $ H_T $ and $ H_U $, respectively, which do not commute, we perform Trotter decomposition to discretize the projector length $ \beta $ into $ L_\tau$ imaginary time slices $ \beta = L_\tau \Delta\tau $. This gives us:
\begin{equation}
\langle \Psi_T \vert e^{-\beta {H}} \vert \Psi_T \rangle = \langle \Psi_T \vert \left(e^{-\Delta_\tau H_U} e^{-\Delta_\tau H_T}\right)^{L_\tau} \vert \Psi_T \rangle + \mathcal{O}\left(\Delta_\tau^2\right), 
\end{equation}
where the non-interacting and interacting parts of the Hamiltonian are separated. The Trotter decomposition introduces a small systematic error of $\mathcal{O}(\Delta\tau^2)$, so we need to set $\Delta\tau$ as a small number to achieve accurate results.

$H_U$ contains a quartic fermionic operator that cannot be measured directly in DQMC. To address this, we employ an SU(2) symmetric Hubbard-Stratonovich (HS) decomposition, where the auxiliary fields couple to the charge density. In our paper, the HS decomposition is as follows:
\begin{equation}
  e^{-\Delta\tau \frac{U}{2}(n_{i,\uparrow}+n_{i,\downarrow}-1)^{2}}=\frac{1}{4}\sum_{\{s_i\}}\gamma(s_{i})e^{\alpha\eta(s_{i})\left(n_{i,\uparrow}+n_{i,\downarrow}-1\right)}+\mathcal{O}\left(\Delta_\tau^4\right),
\label{eq:decompo}
\end{equation}
with $\alpha=\sqrt{-\Delta\tau U/2}$, $\gamma(\pm1)=1+\sqrt{6}/3$,
$\gamma(\pm2)=1-\sqrt{6}/3$, $\eta(\pm1)=\pm\sqrt{2(3-\sqrt{6})}$,
$\eta(\pm2)=\pm\sqrt{2(3+\sqrt{6})}$.

Now, the interacting part is transformed into a quadratic term but is coupled with an auxiliary fields $\{s_{i,l_\tau}\}$ at imaginary time slice $l_\tau$. The subsequent simulations are based on the single-particle basis $\boldsymbol{c} = {c_1, c_2, \cdots,c_{i}, \cdots, c_{n}}$, where $n$ is the system size. Then, we have
   \begin{eqnarray}
\langle\Psi_{T}|e^{-\beta H}|\Psi_{T}\rangle=\sum_{\{s_{i,l_\tau}\}}\left[\left(\prod_{l_\tau}^{L_{\tau}}\prod_{i}^{n}\gamma(s_{i,l_\tau})e^{\alpha\eta(-s_{i,l_\tau})}\right)\det\left[P^{\dagger}B^{s}(\beta,0)P\right]\right]
\label{eq:mcweight}
   \end{eqnarray}
where $P$ is the coefficient matrix of trial wave function $|\Psi_T\rangle$; $B^{s}(\beta,0)$ is defined as
\begin{equation}
B^{s}(\tau_{2},\tau_{1})= \prod_{l_\tau=l_{1}+1}^{l_{2}} \left( \mathrm{e}^{-\Delta_\tau H_T} \prod_{i}^{n} \mathrm{e}^{\alpha\eta(s_{i,l_\tau}) n_{i}} \right)
\end{equation}
with $l_1 \Delta_\tau = \tau_1$ and $l_2 \Delta_\tau = \tau_2$, and has a property $B^{s}(\tau_3,\tau_1)=B^{s}(\tau_3,\tau_2)B^{s}(\tau_2,\tau_1)$.
We could then express the evaluation of the physical operator $\hat{O}$ using the projector DQMC method in a more refined manner as follows:
\begin{equation}\label{eq:s1}
\langle \hat{O} \rangle = \frac{\sum_{\left\{s_{i,l_\tau}\right\}} W_{s_{i,l_\tau}} O_{s_{i,l_\tau}} }{\sum_{\left\{s_{i,l_\tau}\right\}} W_{s_{i,l_\tau}}},
\end{equation}
where $W_{s_{i,l_\tau}} = \left(\prod_{l_\tau}^{L_{\tau}}\prod_{i}^{n}\gamma(s_{i,l_\tau})e^{\alpha\eta(-s_{i,l_\tau})}\right)\det\left[P^{\dagger}B^{s}(\beta,0)P\right] $. 

We further introduce the notation $B^{\rangle}_\tau = B^{s}(\tau,0)P$ and $B^{\langle}_\tau =P^\dagger B^{s}(\beta,\tau)$.
The equal time Green's functions are given as
\begin{equation}
G^{s}(\tau)=\mathbb{I}-B^{\rangle}_\tau \left(B^{\langle}_\tau B^{\rangle}_\tau \right)^{-1}B^{\langle}_\tau,
\end{equation}
where $\mathbb{I}$ is the identity matrix. On a finite precision computer, repeated multiplications between matrices with exponentially large and small values would lead to serious numerical instabilities when straightforwardly calculating $B^{\rangle}_\tau$ and $B^{\langle}_\tau$. To circumvent this problem, UDV matrix decompositions are introduced in the projector DQMC method~\cite{assaadWorld-line2008}.
In practice, we decompose $B^{\rangle}_{\tau}$ into $B^{\rangle}_{\tau} = U^{\rangle}_{\tau} D^{R}_{\tau} V^{R}_{\tau}$ and, similarly, decompose $B^{\langle}_\tau$ into $B^{\langle}_\tau = V^{L}_{\tau} D^{L}_{\tau} U^{\langle}_{\tau}$. This approach allows us to compute the equal-time Green's functions more accurately.
\begin{equation}\begin{aligned}
G(\tau)&=\mathbb{I}-B^{\rangle}_\tau \left(B^{\langle}_\tau B^{\rangle}_\tau \right)^{-1}B^{\langle}_\tau \\
 &= \mathbb{I}-U^{\rangle}_\tau \left(U^{\langle}_\tau U^{\rangle}_\tau \right)^{-1}U^{\langle}_\tau,
\label{eq:ns}
\end{aligned}\end{equation}
because the $D$ and $V$ matrices, which contain extremely large and small eigenvalues, are canceled after every numerical stabilization procedure.

Other physical observables can be calculated from the single-particle Green function using Wick's theorem. More technical details of the projector DQMC algorithms can be found in Refs.~\cite{assaadWorld-line2008}.

From here, for convenience, we denote Eq.~\eqref{eq:s1} as 
\begin{equation}
\langle \hat{O} \rangle = \frac{\sum_{\left\{s\right\}} W_{s} O_{s} }{\sum_{\left\{s\right\}} W_{s}}.
\end{equation}
Grover's pioneering work~\cite{groverEntanglement2013} introduced a general method that involves using two independent replica configurations, labeled as 1 and 2, of the auxiliary field to calculate the 2nd R\'enyi EE for an interacting fermionic system within the DQMC framework,
\begin{equation}
e^{-S_2^A}=\frac{\sum_{\left\{s_1\right\},\left\{s_2\right\}} W_{s_1}W_{s_2} \operatorname{det} g_A^{s_1, s_2}}{\sum_{\left\{s_1\right\},\left\{s_2\right\}} W_{s_1}W_{s_2}},
\end{equation}
where Grover matrix, defined as $ g_A^{s_1,s_2}= G_A^{s_1} G_A^{s_2} + \left(\mathbb{I} - G_A^{s_1}\right)\left(\mathbb{I} - G_A^{s_2}\right) $, is a functional of the Green's functions $ G_A^{s_1} $ and $ G_A^{s_2} $ of two independent replicas. In the main text, we use the notation $ W_{s_1,s_2} = W_{s_1}W_{s_2} $ to make the equation more concise.

\section{The direct method for calculating $S_u$}
The universal term of EE is defined as the difference between the 2nd R\'enyi EEs for two distinct entanglement regions $A$ and $B$,
\begin{equation}
 S_u = S_2^{A}-S_2^{B}.
\end{equation}

From Grover's paper~\cite{groverEntanglement2013}, we know that the numerical definition of the 2nd R\'enyi EE with regions $A$ and $B$, respectively, for interacting fermions can be explicitly expressed as:
\begin{equation}
e^{-S_2^{A}}=\frac{\sum_{\left\{s_1\right\},\left\{s_2\right\}} W_{s_1,s_2} \operatorname{det} g_{A}^{s_1, s_2}}{\sum_{\left\{s_1\right\},\left\{s_2\right\}} W_{s_1,s_2}},
\end{equation}
\begin{equation}
e^{-S_2^{B}}=\frac{\sum_{\left\{s_1\right\},\left\{s_2\right\}} W_{s_1,s_2} \operatorname{det} g_{B}^{s_1, s_2}}{\sum_{\left\{s_1\right\},\left\{s_2\right\}} W_{s_1,s_2}}.
\end{equation}
Thus 
\begin{equation}
\begin{aligned}
e^{-S_u} & \equiv e^{-\left( S_2^{A}-S_2^{B} \right)} \\
& =\frac{\sum_{\left\{s_1\right\},\left\{s_2\right\}} W_{s_1,s_2} \operatorname{det} g_{A}^{s_1, s_2}}{\sum_{\left\{s_1\right\},\left\{s_2\right\}} W_{s_1,s_2}} / \frac{\sum_{\left\{s_1\right\},\left\{s_2\right\}} W_{s_1,s_2} \operatorname{det} g_{B}^{s_1, s_2}}{\sum_{\left\{s_1\right\},\left\{s_2\right\}} W_{s_1,s_2}} \\
& = \frac{\sum_{\left\{s_1\right\},\left\{s_2\right\}} W_{s_1,s_2} \operatorname{det} g_{A}^{s_1, s_2}}{\sum_{\left\{s_1\right\},\left\{s_2\right\}} W_{s_1,s_2} \operatorname{det} g_{B}^{s_1, s_2}}\\
& = \frac{\sum_{\left\{s_1\right\},\left\{s_2\right\}} W_{s_1,s_2} \operatorname{det} g_{B}^{s_1, s_2} \frac{\operatorname{det} g_{A}^{s_1, s_2}}{\operatorname{det} g_{B}^{s_1, s_2}} }{\sum_{\left\{s_1\right\},\left\{s_2\right\}} W_{s_1,s_2} \operatorname{det} g_{B}^{s_1, s_2}}
\end{aligned}
\end{equation}
We use $\bm{W}=  W_{s_1,s_2} \operatorname{det} g_{B}^{s_1, s_2}$ and $\bm{O} = {\operatorname{det} g_{A}^{s_1, s_2}}/{\operatorname{det} g_{B}^{s_1, s_2}} $ to denote the new updating weight and new sampling observable, respectively.
We could calculate $S_u$ via DQCM according to 
\begin{equation}\label{eq:Seq5}
e^{-S_u}=\frac{\sum \bm{W} \bm{O} }{\sum \bm{W}}.
\end{equation}
Using Eq.~\eqref{eq:Seq5}, we can directly calculate the universal term of EE by choosing suitable entanglement regions $A$ and $B$ with the same boundary length, the leading term of the EE can be eliminated. This leaves only the universal sub-leading term, thereby making the extraction of universal features from data fitting more convenient and reliable.

\section{$S_u$ is an exponential observable}
\begin{figure}[htp!]
\includegraphics[width=0.45\columnwidth]{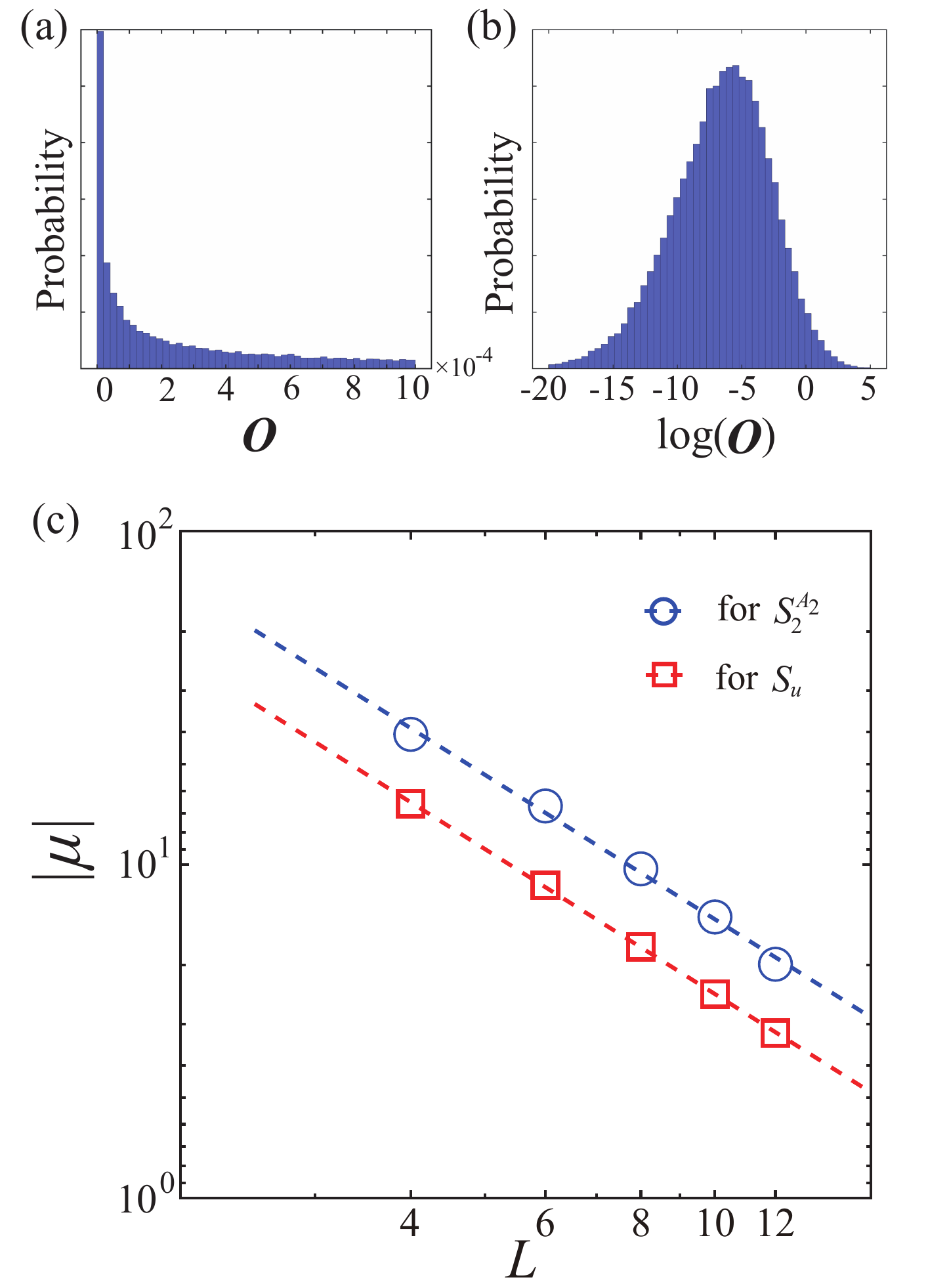}
\caption{The distribution of  $\bm{O}$ (a) and $\ln \bm{O}$ (b) for the $\pi$-flux Hubbard model at $U_c$. (c) The statical mean of $\ln \bm{O}$ for $S_u$ and $\ln (\det g)$ for $A_2$.}
\label{fig:S1}
\end{figure}

As depicted in Fig.~\ref{fig:S1}(a), the distribution of $\bm{O}$ follows a log-normal distribution when we simply use Eq.~\eqref{eq:Seq5} to compute $S_u$ for the $\pi$-flux Hubbard model at $U_c$ with entanglement regions $A_2$ and $A_3$, Here $A_2$ and $A_3$ is given in the main text. Meanwhile, the distribution of $\ln \bm{O}$ conforms well to the normal distribution $\mathcal{N}(\mu,\sigma)$, as illustrated in Fig.~\ref{fig:S1}(b). These observations indicate that, similar to the exponential observable 2nd R\'enyi EE $S_2^A$, $S_u$ is also an exponential observable. 
In other words, we need to employ the power incremental method to calculate $S_u$ rather than simply simulating it according to Eq.~\eqref{eq:Seq5}.

As discussed in the main text, the total incremental steps $N$ can be set as the ceiling number of $|\mu|$. As shown in Fig.~\ref{fig:S1}(c), we observe that the magnitude of $\mu$ of $\ln \bm{O}$ for $S_u$ is smaller than that of $\ln (\det g)$ for $S_2^{A_2}$. This observation implies that we can allocate less CPU time to obtain $S_u$ and subsequently extract $b$ compared to the time required for obtaining $S_2^{A_2}$.

\section{The finite-size effect of $S_u$ at $U=0$}
For the $\pi$-flux Hubbard model at $U=0$, we can obtain the single-particle Green function directly and easily compute the universal term of  EE $S_u$. Here, $S_u$ represents the subtraction between entanglement regions $A_2$ and $A_3$, which implies
\begin{equation}
S_u = b(\pi/2) \ln L+c+\mathcal{O}(1/L).
\end{equation}

\begin{figure}[htp!]
\includegraphics[width=.5\columnwidth]{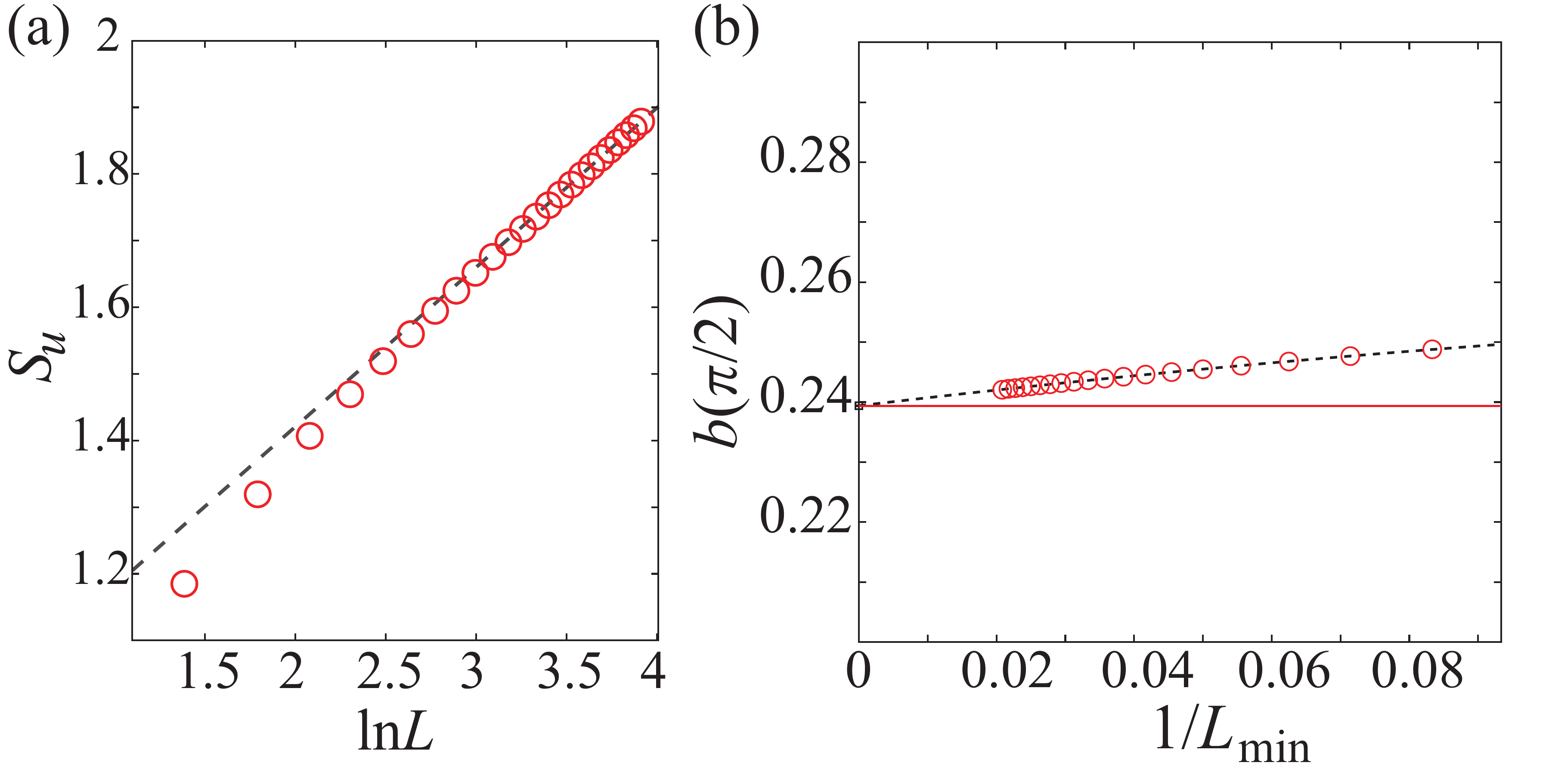}
\caption{(a) $S_u$ as function of $\ln L$, the slope of the dashed line here is $0.23936$. (b) The $1/L_{\text{min}}$ extrapolation of $b(\pi/2)$ to thermodynamic limit yields an extrapolated value consistent with previous predictions of around $0.23936$ (represented by the solid red line).}
\label{fig:S2}
\end{figure}

We expect the universal coefficient $b(\pi/2)$ flux to be $b(\pi/2) = 0.23936$\cite{helmesUniversal2016}. However, as shown in Fig.\ref{fig:S2}(a), when we plot the universal term $S_u$ as a function of $\ln L$, where $L$ is the system size, we observe that $S_u$ exhibits an improved linear relationship with $\ln L$ as $L$ increases.
This phenomenon is reasonable due to the presence of a correction term of order $\mathcal{O}(1/L)$. The slope of the dashed line in Fig.\ref{fig:S2}(a) corresponds to the expected value of $b(\pi/2) = 0.23936$. For smaller system sizes, we cannot obtain an exact value consistent with previous predictions because the correction $\mathcal{O}(1/L)$ becomes more significant, leading to deviations from the expected universal scaling behavior.
As the system size $L$ increases, the contribution from the correction term $\mathcal{O}(1/L)$ diminishes, and the universal scaling of $S_u$ with $\ln L$ becomes more apparent. 

We can extrapolate the extracted value of $b(\pi/2)$ with fitting window $[L_\text{min},50]$, as shown in Fig.~\ref{fig:S2}(b). Consequently, our numerical results approach the predicted value of $b(\pi/2) = 0.23936$ at thermodynamic limit.

\section{Detailed comparison of Method 1, Method 2 and our power incremental method}
In the main text, we demonstrate that our power incremental method outperforms both Method 1, developed by Grover~\cite{groverEntanglement2013}, and Method 2, developed by D'Emidio et al.~\cite{demidioUniversal2022}.
This section provides a more detailed discussion of the advantages of our method.

\subsection{The discussion about the total computational complexity for $S_2^A$ with different methods}
Deterministic Quantum Monte Carlo (DQMC) is an unbiased method that successfully reduces computational complexity from exponential to polynomial scale through importance sampling. It provides a reliable estimation $\langle O \rangle$ of a physical observable $O$ when the variance $\sigma_{O}^2$ is finite, as guaranteed by the central limit theorem.

Our goal is to minimize the relative error $\frac{\sigma_{O}}{\sqrt{n}\langle O \rangle} \equiv \text{CV}[O]/\sqrt{n}$, which requires sampling as many n samples as possible. However, we are limited to a polynomial scale of samples in practice simulation. As shown in our main text, the coefficient of variation (CV) of $(\det g)$ grows exponentially with increasing $L$, causing the total computational complexity for Method 1 to grow exponentially and become unfeasible. Consequently, Method 1 fails to provide reliable estimates of $(\det g)$ for relatively large $L$. Our numerical results, presented in Fig.4(a) of the main text, clearly illustrate this challenge.

Fortunately, we observe that the CV of $(\det g)^{1/N}$ decreases with $L$, allowing us to obtain the mean of $(\det g)^{1/N}$, i.e., Z(k+1)/Z(k), with polynomial-growing effort. Here, $N \sim L^{\gamma}$ is the ceiling integer of $|\langle \log(\det g) \rangle|$. The overall computational complexity for computing $S_2^A(L)$ is $O(\beta L^{6+\gamma} )$.

It is crucial to note that our power incremental method successfully reduces the total computational burden from exponential to polynomial growth as $L$ increases, in contrast to Method 1. This improvement gives us confidence in our method's ability to reliably estimate the entanglement entropy (EE).

Method 2 also could reduce the total computational burden from exponential to polynomial growth as $L$ increases with help the similar incremental procedure. However, Method 2 introduces an additional configuration space, leading to an increased computational burden for calculating Z(k+1)/Z(k) than our method. More critically, the estimator utilized in Ref.\cite{demidioUniversal2022} could not revert to the original definition of EE~\cite{groverEntanglement2013} when the total number of incremental processes was reduced to 1, as discussed below.
Consequently, the total number of incremental processes for Method 2 generally needs to be quite large, and of course Method 2 have worse efficient than our method.

\subsection{More details of Method 2 }
The detailed methodology of Method 2, as presented in Ref~\cite{demidioUniversal2022}, is as follows.
They define
\begin{equation}
Z(\lambda_k) = \sum_{{s_1, s_2, C \subseteq A}} \lambda_k^{N_C}(1-\lambda_k)^{N_A-N_C} W_{s_1} W_{s_2} \operatorname{det} g_C^{s_1, s_2},
\end{equation}
where $C$ represents proper subsets of the entanglement region $A$, and $N_C$ ($N_A$) is the number of sites in region $C$ ($A$), $\lambda_k$ is a real number between 0 and 1.
They then calculate the EE in an incremental form:
\begin{equation}\label{eq:incremental}
e^{-S_2^A} = \frac{Z(\lambda_1)}{Z(\lambda_0)}\frac{Z(\lambda_2)}{Z(\lambda_1)}\cdots\frac{Z(\lambda_{k+1})}{Z(\lambda_k)} \cdots\frac{Z(\lambda_{N})}{Z(\lambda_{N - 1})},
\end{equation}
where $k$ represents the $k$-th incremental process, with a total of $N$ incremental processes.
They will have $Z(\lambda_0 = 0) = \sum_{{s_1},{s_2}} W_{s_1} W_{s_2}$ and $Z(\lambda_{N} = 1) = \sum_{{s_1},{s_2}} W_{s_1} W_{s_2} \operatorname{det} g_A^{s_1, s_2}$. This definition appears to satisfy the requirement of Method 1, i.e., $e^{-S_2^A} = \langle \operatorname{det} g_A^{s_1, s_2} \rangle$.
However, in practice, they simulate $\frac{Z(\lambda_{k+1})}{Z(\lambda_k)}$ as:
\begin{equation}
\frac{Z(\lambda_{k+1})}{Z(\lambda_k)} = \frac{\sum_{{s_1, s_2, C \subseteq A}} \mathbf{W}^{s_1,s_2}_{C,\lambda_k} O_{\lambda_k,\lambda_{k+1}} }{\sum_{{s_1, s_2, C \subseteq A}} \mathbf{W}^{s_1,s_2}_{C,\lambda_k} },
\end{equation}
where the modified sampling weight is
$
\mathbf{W}^{s_1,s2}_{C,\lambda_k} = \lambda_k^{N_C}(1-\lambda_k)^{N_A-N_C} W{s_1} W{s_2} \operatorname{det} g_C^{s_1, s_2},
$
and the modified sampling observable is $O_{\lambda_k,\lambda_{k+1}} = \left(\frac{\lambda_{k+1}}{\lambda_k}\right)^{N_C}\left(\frac{1-\lambda_{k+1}}{1-\lambda_k}\right)^{N_A-N_C}$.
Notably, if there is only one incremental step (i.e., $N=1$), Method 2 is not equivalent to Method 1. According to Method 2:
\begin{equation}
e^{-S_2^A} = \frac{Z(1)}{Z(0)} = \frac{\sum_{{s_1, s_2, C \subseteq A}} 0^{N_C} 1^{N_A-N_C} W_{s_1} W_{s_2} \operatorname{det} g_C^{s_1, s_2} \left(\frac{1}{0}\right)^{N_C}\left(\frac{0}{1}\right)^{N_A-N_C} }{\sum_{{s_1, s_2, C \subseteq A}} 0^{N_C} 1^{N_A-N_C} W_{s_1} W_{s_2} \operatorname{det} g_C^{s_1, s_2}},
\end{equation}
which implies that Method 2 is used to sample $(1/0)^{N_C}(0/1)^{N_A-N_C}$. Given the updating strategy of Method 2, if $\lambda_k=0$, then $N_C$ must be 0, leading to the value of all samples $(1/0)^{N_C}(0/1)^{N_A-N_C} = 0$. Consequently, we get $S^A_2 = -\ln(0) = \infty$. This significant bias explains why Method 2 performs much worse than Method 1 when $N$ is very small.
In contrast, our method yields:
\begin{equation}\label{eq:4}
\frac{Z({k+1})}{Z(k)} = \frac{\sum_{{s_1},{s_2}} W_{s_1,s_2} (\operatorname{det} g_A^{s_1, s_2})^{k/ N} (\operatorname{det} g_A^{s_1, s_2})^{1/N} }{\sum_{{s_1},{s_2}} W_{s_1,s_2} (\operatorname{det} g_A^{s_1, s_2})^{k/N}},
\end{equation}
When $N=1$, our method reduces to $e^{-S_2^A} =\frac{Z(1)}{Z(0)} = \langle \operatorname{det} g_A^{s_1, s_2} \rangle$, exactly the same as Method 1.
This consistency explains why our method provides better results, even for small N.
\begin{figure}
\centering
\includegraphics[width=0.4\columnwidth]{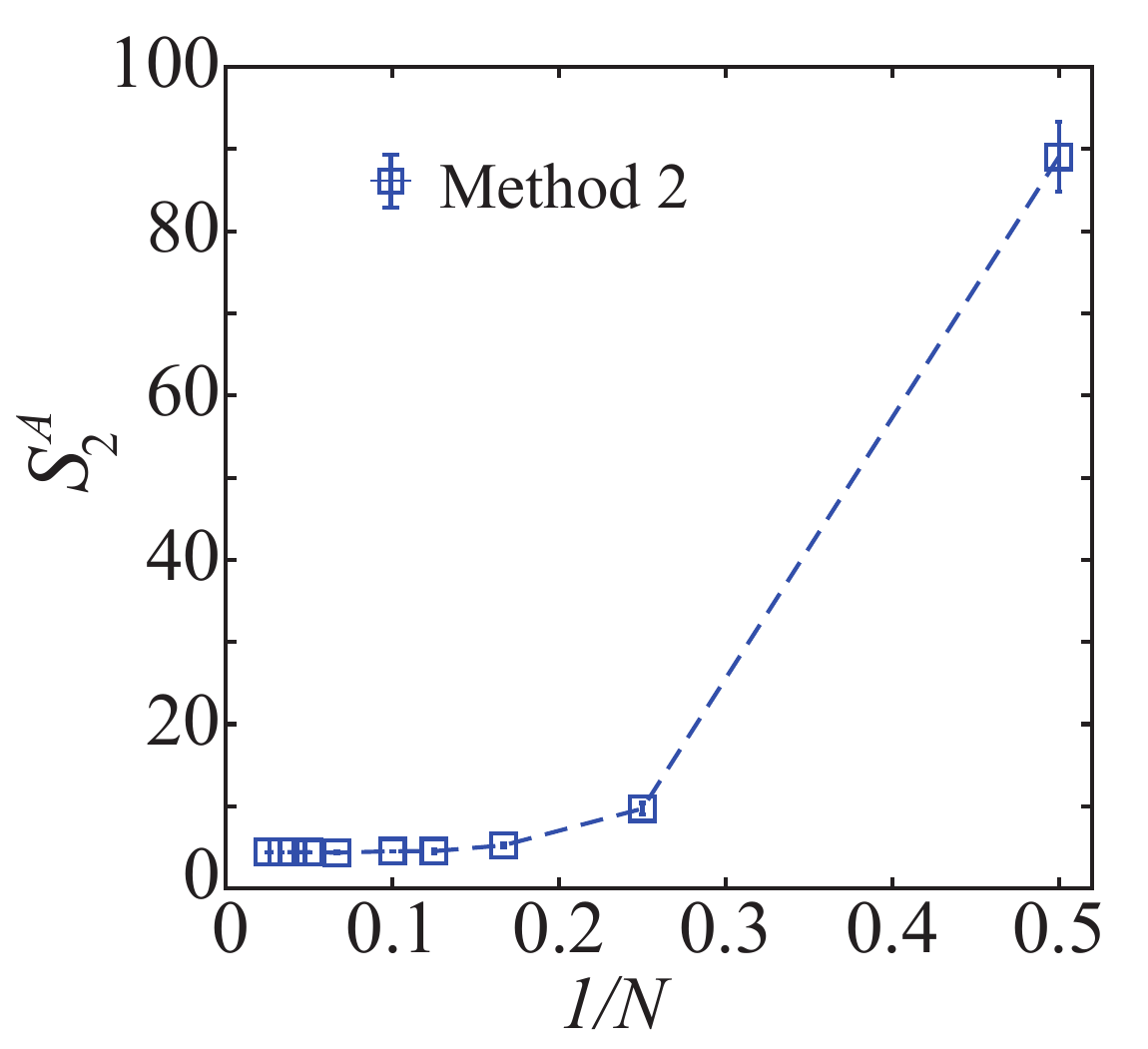}
\caption{The $1/N$ extrapolation of $S^A_2$ for Method 2 from $N=2$ to $N=40$.}
\label{fig:S4}
\end{figure}
Fig.~\ref{fig:S4} shows the $S^A_2$ data for different $N$ values (excluding $N=1$) using Method 2. We exclude $N=1$ because the $S^A_2$ obtained by Method 2 at $N=1$ is $\infty$ and cannot be plotted

\section{Positivity of Grover matrix determinant}
The Grover matrix is defined as 
\begin{equation}
g_A^{s_1,s_2}= G_A^{s_1} G_A^{s_2}+\left(\mathbb{I}-G_A^{s_1}\right)\left(\mathbb{I}-G_A^{s_2}\right),
\end{equation}
where $ G_A^{s_1} $ and $ G_A^{s_2} $ are Green functions for two independent replicas. As discussed in the main text, to obtain accurate EE, we need to calculate it via DQMC.
\begin{equation}
\frac{Z\left({k+1}\right)}{Z\left(k\right)}=\frac{\sum\limits_{\left\{s_1\right\},\left\{s_2\right\}} W_{s_1,s_2}  ( \operatorname{det} g_A^{s_1, s_2})^{k/ N} ( \operatorname{det} g_A^{s_1, s_2})^{1/N} }{\sum\limits_{\left\{s_1\right\},\left\{s_2\right\}} W_{s_1,s_2} ( \operatorname{det} g_A^{s_1, s_2})^{k/N}}.
\end{equation}
Now, it's crucial to ensure that we include the Grover matrix determinant $ \operatorname{det} g_A^{s_1, s_2} $ in the updating weight. Thus, $ \operatorname{det} g_A^{s_1, s_2} $ must be positive to avoid the sign problem in simulations. In this section, we prove that the Grover matrix determinant is real and positive for the 0-flux and $\pi$-flux Hubbard models at charge neutrality.

The 0-flux and $\pi$-flux Hubbard models are defined on bipartite lattices with two sublattices. At charge neutrality, this model has particle-hole (PH) symmetry, and as proved in Ref.~\cite{yuandaliaoCorrelationInduced2021,xuCompeting2021,wangEntanglement2023a}, one can find that this PH symmetry ensures that the single-particle Green function for spin-up and spin-down flavors has the relation $G_{ij}^{\downarrow} = (-1)^{i+j} (\delta_{ij} - G_{ji}^{\uparrow*})$, which can be rewritten in the following matrix form: $ G^{\downarrow} = U^{\dagger} \left( I - (G^{\uparrow})^{\dagger} \right) U $ with $U_{ij} = \delta_{ij}(-1)^i = \delta_{ij}(-1)^j$ being a diagonal unitary matrix.

This condition is sufficient for $\det g_A^{s_1, s_2} \geq 0$, because we can prove that the Grover matrix $g_A^{s_1, s_2} = g_A^{s_1, s_2, \uparrow} \otimes g_A^{s_1, s_2, \downarrow}$ and $\operatorname{det} g_A^{s_1, s_2, \uparrow} = \left( \operatorname{det} g_A^{s_1, s_2, \downarrow} \right)^{*}$. Then, $ \operatorname{det} g_A^{s_1, s_2} = \operatorname{det} g_A^{s_1, s_2, \uparrow} \cdot \operatorname{det} g_A^{s_1, s_2, \downarrow} = \left| \operatorname{det} g_A^{s_1, s_2, \uparrow} \right|^2. $ This ensures that $\operatorname{det} g_A^{s_1, s_2}$ is real and positive, thereby avoiding the sign problem in simulations.

\end{document}